\documentclass[11pt]{article}
\usepackage[]{algorithm2e}
\usepackage{graphicx,wrapfig,lipsum}
\usepackage{graphicx,psfrag,epsf}
\usepackage{amsmath}
\usepackage{enumerate}
\usepackage{amsthm}
\usepackage{bm}
\usepackage{titling}
\usepackage{natbib}

\usepackage{lipsum}

\usepackage{mathtools}
\usepackage{thmtools}
\usepackage{xr}
\usepackage{amssymb}
\usepackage{slashbox}
\usepackage[utf8]{inputenc}
\usepackage[english]{babel}
\usepackage[margin=1in]{geometry}

\newcommand{\blind}{0}

\newtheorem{theorem}{Theorem}[section]
\newtheorem{corollary}{Corollary}[theorem]
\newtheorem{lemma}[theorem]{Lemma}
\newtheorem{prop}[theorem]{Proposition}
\renewcommand{\baselinestretch}{1.5}
\newcommand{\distas}[1]{\mathbin{\overset{#1}{\kern\z@\sim}}}%
\title{Parallelisation of a Common Changepoint Detection Method}
\author{S. O. Tickle, I. A. Eckley, P. Fearnhead, K. Haynes}
\date{\today}

\newcommand*{\QEDB}{\hfill\ensuremath{\square}}

\begin{document}

\def\spacingset#1{\renewcommand{\baselinestretch}%
{#1}\small\normalsize} \spacingset{1}

\if0\blind
{
  \title{\bf Parallelisation of a Common Changepoint Detection Method}
  \author{S. O. Tickle$^1$\thanks{Email: s.tickle1@lancaster.ac.uk, Address: STOR-i Centre for Doctoral Training, Lancaster University, Lancaster LA1 4YR, United Kingdom} , I. A. Eckley$^2$, P. Fearnhead$^2$, K. Haynes$^1$ 
		\hspace{.2cm}\\
    $^1$STOR-i Centre for Doctoral Training, Lancaster University, Lancaster, United Kingdom; \\
    $^2$Department of Mathematics and Statistics, Lancaster University, Lancaster, United Kingdom
}

  \maketitle
} \fi

\if1\blind
{
  \bigskip
  \bigskip
  \bigskip\include{C:/ProgramData/Microsoft/Windows/Start Menu/Programs/MiKTeX 2.9/TeXworks.lnk}

  \begin{center}
    {\LARGE\bf Title}
\end{center}
  \medskip
} \fi

\begin{abstract}
In recent years, various means of efficiently detecting changepoints in the univariate setting have been proposed, with one popular approach involving minimising a penalised cost function using dynamic programming. In some situations, these algorithms can have an expected computational cost that is linear in the number of data points;  however, the worst case cost remains quadratic. We introduce two means of improving the computational performance of these methods, both based on parallelising the dynamic programming approach. We establish that parallelisation can give substantial computational improvements: in some situations the computational cost decreases roughly quadratically in the number of cores used. These parallel implementations are no longer guaranteed to find the true minimum of the penalised cost; however, we show that they retain the same asymptotic guarantees in terms of their accuracy in estimating the number and location of the changes.
\end{abstract}

\noindent%

\spacingset{1.45}
\section{Introduction}\label{sec:intro}

The challenge of changepoint detection has received considerable interest in recent years (see, for example, \cite{Rigaill12}, \cite{ChenNkurunziza17} and \cite{Truong18} and references therein). In particular, there has been a significant focus on the important issue of developing computationally efficient methods to detect multiple changes. This article makes a new contribution to this area by focusing on the problem of parallelising a penalised cost approach to provide a significant computational advantage without compromising on statistical efficiency.

The common changepoint problem setting considers the analysis of a data sequence, $y_1, ..., y_n$, which is ordered by some index, such as time or position along a chromosome. We use the notation $y_{s:t} = \left(y_s, \ldots, y_t\right)$ for $t \geq s$. Our interest is in segmenting the data into consecutive regions; such a segmentation can be defined by the changepoints, $0 = \tau_0 < \tau_1 < \ldots < \tau_m < \tau_{m+1} = n$, where throughout we take $m$ as fixed, but unknown. Thus the set of changepoints splits the data into $m+1$ segments, with the $j^{th}$ segment containing data-points $y_{\tau_{j-1} + 1:\tau_j}$.

Several approaches can be used to identify the locations of these changes. Within this article, we focus on a class of methods which involve finding the set of changepoints that minimise a given cost. The cost associated with a specific segmentation consists of two important specifications. The first of these is $\mathcal{C}(.)$, the cost incurred from a segment of the data. Common choices for $\mathcal{C}(.)$ include quadratic error loss, Huber loss and the negative log-likelihood (for an appropriate within-segment model for the data); see \cite{YaoAu89}, \cite{FearnheadRigaill17} and \cite{ChenGupta00} for further discussion. For example, using quadratic error loss gives:

\begin{equation}\label{leastsquares}
	\mathcal{C}(y_{s:t}) = \sum_{i=s}^t \left(y_i - \frac{1}{t-s+1} \sum_{j=s}^t y_j \right)^2.
\end{equation}

Note that in the case of a piecewise constant signal observed with additive Gaussian noise,~(\ref{leastsquares}) is equivalent to twice the negative log-likelihood. The second specification is $\beta$, the penalty incurred when introducing a changepoint into the model. Common choices for $\beta$ include the Akaike Information Criterion, Schwarz Information Criterion and modified Bayesian Information Criterion; see \cite{Hocking13}, \cite{HaynesEckleyFearnhead17} and \cite{Truong17} and references therein for further discussion. Finally, it is assumed that the cost function is additive over segments. The objective is then to find the segmentation which minimises the cost. In other words, we wish to find:

\begin{equation}\label{fullcost}
	\arg \min_{1 \leq \tau_1 < \ldots < \tau_m \leq n-1} \sum\limits_{i = 1}^{m+1} \left[\mathcal{C}(y_{\tau_{i-1}+1:\tau_i}) + \beta\right].
\end{equation}

\par Dynamic programming methods exist which are guaranteed to find the global minimum of~(\ref{fullcost}). Optimal Partitioning, due to \cite{JacksonScargleBarnes05}, uses dynamic programming to solve~(\ref{fullcost}) exactly in a computation time of $\mathcal{O}(n^2)$. \cite{KillickFearnheadEckley12} introduce the PELT algorithm, which also solves~(\ref{fullcost}) exactly, and can have a substantially reduced computational cost. In situations where the number of changepoints increases linearly with $n$, \cite{KillickFearnheadEckley12} show that PELT's expected computational cost can be linear in $n$. However, the worst case cost is still $\mathcal{O}(n^2)$, suggesting that significant computational savings are still desirable in practice.

Parallel computing techniques are an increasingly popular means of doing precisely this. The application of parallelisation is vast, with use in such areas as meta-heuristics, cloud computing and biomolecular simulation, as discussed in \cite{Alba05}, \cite{Mezmaz11}, \cite{Schmid12} and \cite{WangDunson14} among many others. Some methods are more easily parallelisable in that it is plain how to split a search space or other task between different nodes. These problems are often described as `Embarrassingly Parallel'. For the changepoint detection problem, some existing methods may be described as such. These include Binary Segmentation, due to \cite{ScottKnott74}, and its related approaches, notably the Wild Binary Segmentation (WBS) method of \cite{Fryzlewicz14}. However, it is not so straightforward to parallelise dynamic programming methods such as PELT. This shall be the focus of this paper.

One of our approaches to parallelising algorithms such as PELT will use the fact that~(\ref{fullcost}) can still be solved exactly when we restrict the changepoints to an ordered subset $\mathcal{B} = \{b_1, \ldots, b_k\} \subset \{ 1, \ldots, \left(n - 1\right) \}$. Let $F(u)$ denote the minimum of~(\ref{fullcost}) when we restrict changepoints to $\mathcal{B}$ and consider data only up to time $u$; in addition let $\left(\tau_1, \ldots, \tau_{m'}\right)$ be an ordered set of (estimated) changepoints, so that, for $t < s$:

\begin{align*}
	F(b_s) &= \min_{m', \hspace{1pt} \left(\tau_1, \ldots, \tau_{m'}\right) \subseteq \{b_1, \ldots, b_{s-1}\}} \sum\limits_{i = 1}^{m'+1} \left[\mathcal{C}(x_{\tau_{i-1}+1:\tau_i}) + \beta\right], \\
	&= \min_t \left\{ \min_{m', \hspace{1pt} \left(\tau_1, \ldots, \tau_{m'}\right) \subseteq \{b_1, \ldots, b_{t-1}\}} \sum\limits_{i = 1}^{m'+1} \left[\mathcal{C}(x_{\tau_{i-1}+1:\tau_i}) + \beta\right] + \mathcal{C}(x_{b_t+1:b_s}) + \beta \right\}, \\
	&= \min_t \left\{ F(b_t) + \mathcal{C}(x_{b_t+1:b_s}) + \beta \right\}.
\end{align*}

Using the initial condition $F(0) = 0$, this gives a means of recursively calculating $F(b_k)$. 

The general format of this paper is as follows: Section~\ref{sec:parallelisation} introduces two means of parallelising dynamic programming methods for solving (\ref{fullcost}), which we refer to as \textbf{Chunk} and \textbf{Deal}. In each case, we provide a description of the proposed algorithm with practical suggestions for implementation, followed by a short discussion of the theoretical justifications behind these choices. We devote Section~\ref{sec:theory} to examining this latter aspect in detail. In particular, we establish the asymptotic consistency of Chunk and Deal in a specific case with recourse to the asymptotic consistency of the penalised cost function method. Section~\ref{sec:sims} compares the use of parallelisation to other common approaches in a number of scenarios involving changes in mean. We conclude with a short discussion in Section~\ref{sec:discussion}. The proofs of all results may be found in the appendices and supplementary materials.

\section{Parallelisation of Dynamic Programming Methods}\label{sec:parallelisation}

In this section, we introduce Chunk and Deal, two methods for parallelising dynamic programming procedures for changepoint detection. For convenience, we shall herein refer to this exclusively as the parallelisation of PELT.

We introduce the notation $PELT(y_\mathcal{A}, \mathcal{B})$ when referring to applying PELT to a dataset $y_\mathcal{A}$ but only allowing candidate changepoints to be fitted from within the set $\mathcal{B}$. Note that we trivially require $\mathcal{B} \subseteq \mathcal{A}$. The general setup for the parallelisation procedure then takes the following form:

\begin{itemize}
\item (Split Phase) We divide the space $\left\{1, \ldots, \left(n - 1\right) \right\}$ into (not necessarily disjoint) subsets $\mathcal{B}_1, \ldots, \mathcal{B}_{L\left(n\right)}$, where $L(n)$ is the number of computer cores available;
\item Each of the cores $i = 1, \ldots, L(n)$ then performs $PELT(y_{\mathcal{A}_i}, \mathcal{B}_i)$, returning a candidate set, $\boldsymbol{\hat{\tau}}_i$, of changes, which are returned to the parent core;
\item (Merge Phase) The parent core then performs $PELT(y_{1:n}, \cup_{i=1}^{L\left(n\right)} \boldsymbol{\hat{\tau}}_i)$, and the method returns $\boldsymbol{\hat{\tau}}$, the set of estimated changes found at this stage.
\end{itemize}

Note that in the above we require $\cup_{i=1}^{L\left(n\right)} \mathcal{A}_i = \left\{1, \ldots, n \right\}$.

\subsection{Chunk}\label{ssec:chunk}

\par

\begin{figure}[h!]
\begin{minipage}[t]{0.46\textwidth}
\includegraphics[width=\linewidth,keepaspectratio=true]{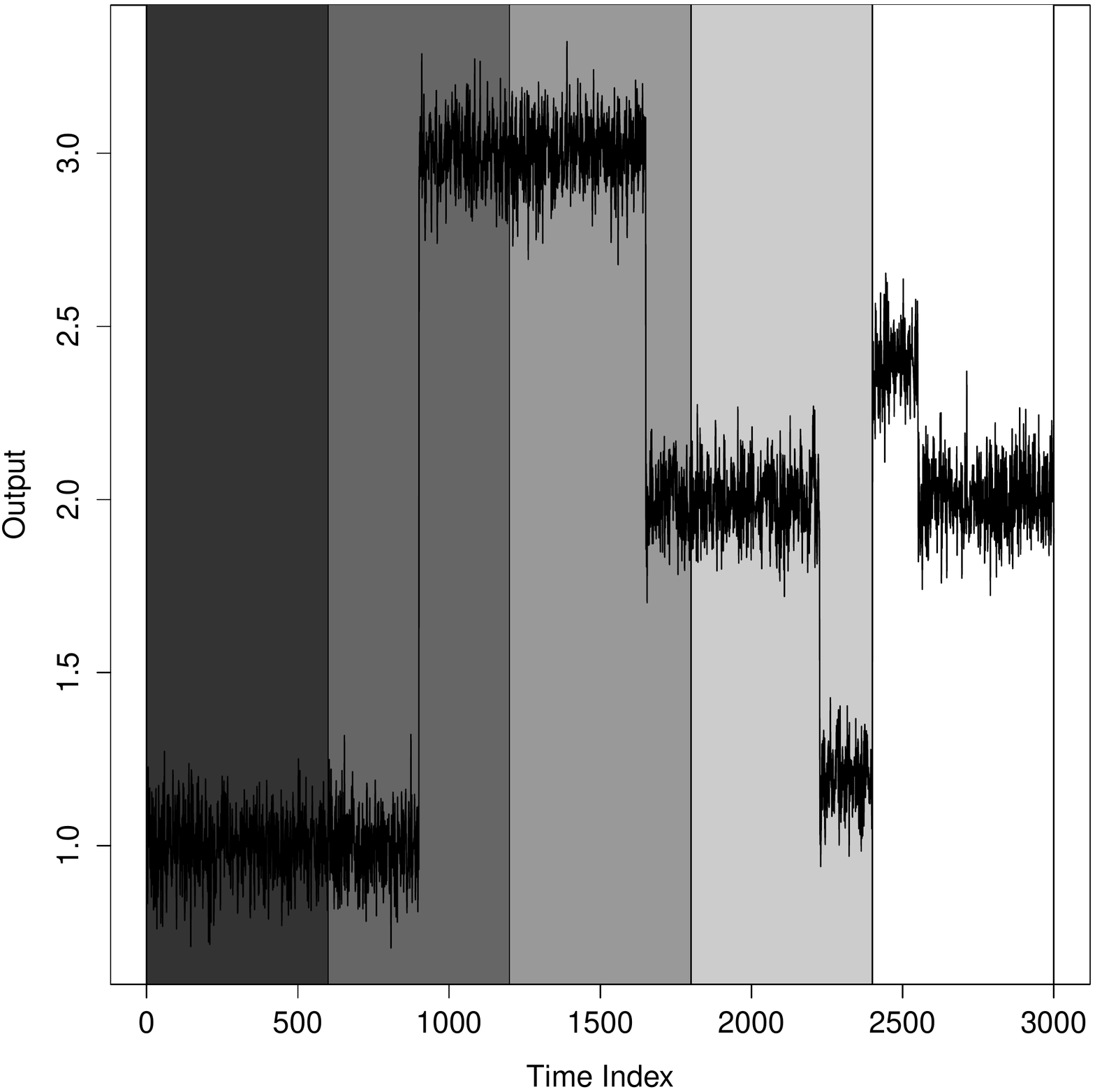}
\end{minipage}
\hspace*{\fill} 
\begin{minipage}[t]{0.46\textwidth}
\includegraphics[width=\linewidth,keepaspectratio=true]{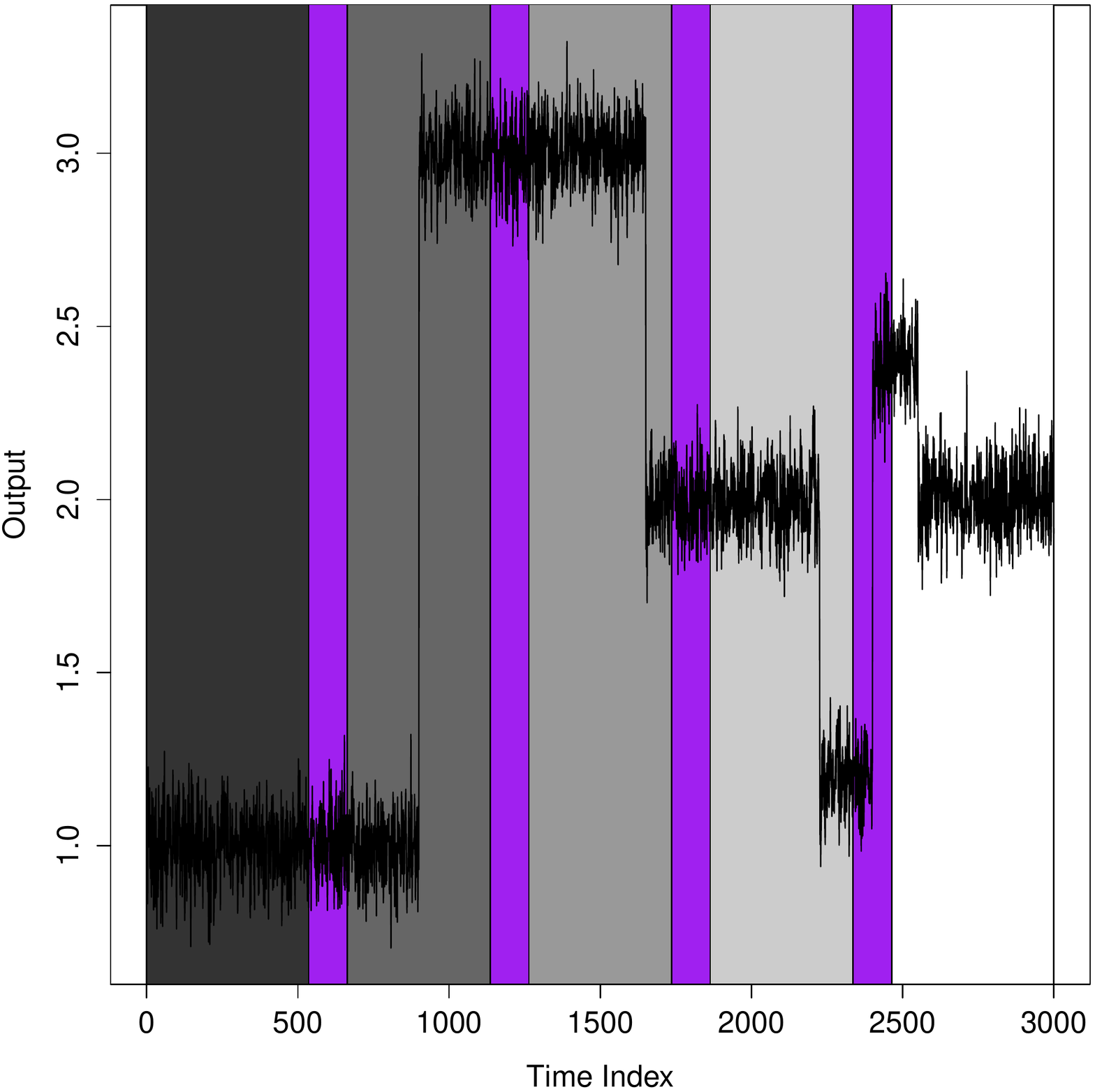}
\end{minipage}
\caption{The time series is split into continuous segments by the Chunk procedure, in this case with 5 cores (l). An overlap is specified between the segments such that points within are considered by both adjacent cores (r).}
\label{phase2}
\end{figure}

The Chunk procedure consists of dividing the data into continuous segments and then handing each core a separate segment on which to search for changes. This splitting mechanism is shown in Figure~\ref{phase2}. One problem with this division arises from changes which can be arbitrarily close to, or coincide with, the `boundary points' of adjacent cores. This necessitates the use of an overlap - a set of points which are considered by both adjacent cores for potential changes, also shown in Figure~\ref{phase2}. For a time series of length $n$, we choose an overlap of size $V(n)$ either side of the boundary for each core. The full procedure for Chunk is detailed in Algorithm 1.

\begin{algorithm}[H]
 \caption{Chunk for the PELT procedure}
 \KwData{A univariate dataset, $y_{1:n}$.  }
 \KwResult{A set of estimated \textbf{changepoint locations} $\hat{\tau}_1, \ldots, \hat{\tau}_{\hat{m}}$.}
 \textbf{Step 1}: Split the dataset into the subsets $\mathcal{B}_1, \ldots, \mathcal{B}_{L(n)}$ such that $\mathcal{B}_1 = \left\{1, \ldots, {\left\lfloor\frac{n}{L\left(n\right)}\right\rfloor + V(n)} \right\}$, $\mathcal{B}_{L\left(n\right)} = \left\{\left(L(n)-1\right)\left\lfloor \frac{n}{L\left(n\right)} \right\rfloor - V(n), \ldots, n \right\}$,  $\mathcal{B}_i = \left\{ \left(i-1\right)\left\lfloor \frac{n}{L\left(n\right)} \right\rfloor - V(n), \ldots, i \left\lfloor \frac{n}{L\left(n\right)} \right\rfloor + V(n) \right\}$ $\forall i \in \left\{2, \ldots, L(n) - 1 \right\}$;

 \For{$i = 1, \ldots, L(n)$}{
  On core $i$, find $\boldsymbol{\hat{\tau}}_i = PELT\left(y_{\mathcal{B}_i}, \mathcal{B}_i \right)$;
 }

 \textbf{Step 2}: Sort $\cup_{i=1}^{L\left(n\right)} \boldsymbol{\hat{\tau}}_i$ into ascending order;

 \textbf{Step 3}: Calculate and return $\left(\hat{\tau}_1, \ldots, \hat{\tau}_{\hat{m}}\right) = PELT\left(y_{1:n}, \cup_{i=1}^{L\left(n\right)} \boldsymbol{\hat{\tau}}_i \right)$.

\vspace{30pt}
\end{algorithm}

Given that Algorithm 1 executes PELT multiple times, it is not immediate that Chunk represents a computational gain. We therefore briefly examine the speed of the procedure from an intuitive perspective. Taking the worst case computational cost of PELT to be $\mathcal{O}(|\mathcal{B}|^2)$, where $\mathcal{B}$ is the candidate set of changepoints, then the worst case cost of the split phase will be $\mathcal{O}\left(\left(\frac{n}{L\left(n\right)}\right)^2\right)$. The cost of the merge phase is dependent on the total number of estimated changes generated in the split phase. If we can estimate changepoint locations to sufficient accuracy, then as each change appears in at most two of the `chunks', the number of returned changes ought to be at most $2m$. Thus the merge phase has a cost that is $\mathcal{O}(m^2)$. This intuition is confirmed later, in Corollary~\ref{parspeed}.

In order to guarantee that the method does not overestimate the number of changes, some knowledge of the location error inherent in the PELT procedure is needed. This motivates the results of Section~\ref{sec:theory}, which in turn imply various practical choices for the length of the overlap region, $V(n)$. In particular, using $V(n) = \left\lceil \left(\log n\right)^2 \right\rceil$ will give an effective guarantee of the accuracy of the method. Other sensible choices for $V(n)$ can be made based on the trade-off between accuracy and speed (see Section~\ref{sec:theory} for details).

\subsection{Deal}\label{ssec:deal}

\par

The Deal procedure relies on distributing points to the computing cores in the same manner as a playing card dealer. Define $Q_a(b,c)$ as the largest integer such that $Q_a(b,c) \times b + \left(a \hspace{-4pt} \mod b\right) < c$. The split phase then partitions $\left\{1, \ldots, \left(n-1\right) \right\}$ as follows:

\begin{align*}
\mathcal{B}_1 &= \{1, {L(n)+1}, {2L(n) + 1}, \ldots, {Q_1(L(n),n) L(n) + 1} \},\\
\mathcal{B}_2 &= \{2, {L(n)+2}, {2L(n) + 2}, \ldots, {Q_2(L(n),n)L(n) + 2} \},\\
&\ldots \\
\mathcal{B}_{L(n)} &= \{L(n), {2L(n)}, {3L(n)}, \ldots, {Q_{L\left(n\right)}(L(n),n)L(n)} \}.
\end{align*}

\begin{figure}[h!]
\begin{minipage}[t]{\textwidth}
\vspace{-10pt}
\begin{center}
\includegraphics[width=0.5\linewidth,keepaspectratio=true]{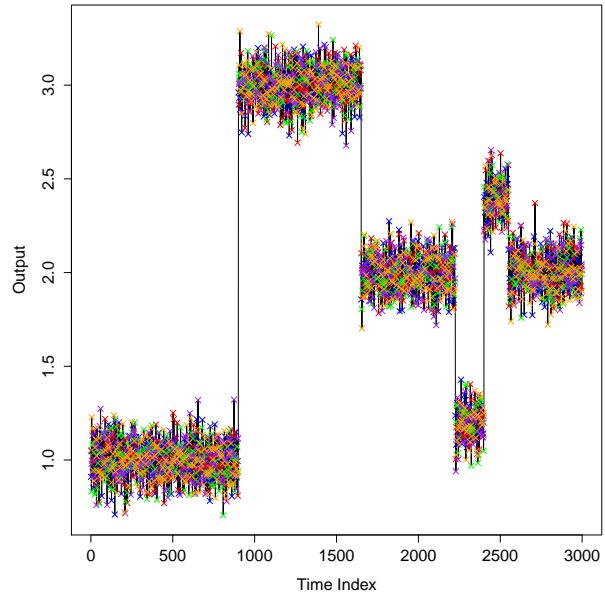}
\end{center}
\vspace{-20pt}
\caption{Points coloured differently are processed by different cores. In the above case 5 cores are used. A core may fit changes only in locations of a given colour.}
\label{phase 3}
\end{minipage}
\end{figure}

This splitting mechanism is shown in Figure~\ref{phase 3}. On the $k^{th}$ core, the objective function to be minimised then becomes:

\begin{center}

$\min\limits_{m, \tau_1, \ldots, \tau_m \in \mathcal{B}_k} \sum\limits_{i=1}^{m+1} \left\{\mathcal{C}(y_{\left(\tau_{i-1} + 1\right):\tau_{i}}) + \beta \right\}$,

\end{center}

as discussed in Section~\ref{sec:intro}. The full procedure for Deal is detailed in Algorithm 2.

\begin{algorithm}[H]
 \caption{Deal for the PELT procedure}
 \KwData{A univariate dataset, $y_{1:n}$.  }
 \KwResult{A set of estimated \textbf{changepoint locations} $\hat{\tau}_1, \ldots, \hat{\tau}_{\hat{m}}$.}
 \textbf{Step 1}: Split the dataset into subsets $\mathcal{B}_1, \ldots, \mathcal{B}_{L(n)}$ such that $\mathcal{B}_i =\left\{i, L(n) + i, \ldots, Q_i(L(n), n) + \left(i \hspace{-4pt} \mod L(n)\right) \right\}$;

 \For{$i = 1, \ldots, L(n)$}{
  On core $i$, find $\boldsymbol{\hat{\tau}}_i = PELT\left(y_{1:n}, {\mathcal{B}_i}\right)$;
 }

 \textbf{Step 2}: Sort $\cup_{i=1}^{L\left(n\right)} \boldsymbol{\hat{\tau}}_i$ into ascending order;

 \textbf{Step 3}: Calculate and return $\left(\hat{\tau}_1, \ldots, \hat{\tau}_{\hat{m}}\right) = PELT\left(y_{1:n}, \cup_{i=1}^{L\left(n\right)} \boldsymbol{\hat{\tau}}_i \right)$.

\vspace{30pt}

\end{algorithm}

As for the Chunk procedure, the implementation of Deal leads to computational gains. By the previous section, the worst case computational time of the split phase of Deal will be $\mathcal{O}\left(\left(\frac{n}{L\left(n\right)}\right)^2\right)$. The speed of the merge phase is again dependent on the number of changes detected at the split phase. We demonstrate in the proof of Corollary~\ref{parspeed} that the number of changes detected by each core is at most $2m$, meaning that the worst case performance of the merge phase is $\mathcal{O}\left(L(n)^2\right)$.

For both procedures, using the standard SIC penalty gives a maximum location error of $\mathcal{O}(\log n)$ in the asymptotic setting, see Theorems~\ref{chunkconsistency} and~\ref{dealconsistency} for details. With the Deal procedure, however, an additional lower bound constraint is enforced on the number of cores required for this location error (see Theorem~\ref{dealconsistency}); we therefore recommend setting $L(n)$ to be as large as the number of cores available in most practical settings.

We remark that while the Chunk and Deal procedures do not inherit the exactness of PELT in finding the optimal solution to~(\ref{fullcost}), they nevertheless track the true optimum very closely, as seen by the empirical results in Section~\ref{sec:sims}.

\section{Consistency of Parallelised Approaches}\label{sec:theory}

As exactness with respect to minimising~(\ref{fullcost}) cannot be assumed for the two methods, we must verify that they retain the desirable properties of PELT. To this end, we now turn to consider the consistency of the parallel procedures for a change in mean setting.

We now stipulate that a time series $y_1, ..., y_n$ has changepoints corresponding to proportions $\theta_1, ..., \theta_m$, for some fixed $m$, such that, for a given $n$, the changepoints $\tau_1, ..., \tau_m$ are defined as $\tau_i = \left\lfloor \theta_i n \right\rfloor$ $\forall i$. For the asymptotic setting we consider, take $\theta_{1:m}$ to be fixed.

With this framework in place, we note that the consistency results for Chunk and Deal we develop in Section \ref{ssec:compcost} require one particular result not provided by \cite{KillickFearnheadEckley12}, namely consistency of PELT for the change in mean setting.

\begin{prop}\label{peltconsistency}
We consider the change in mean setting for the univariate time series: 

\begin{equation}\label{changeinmean}
Y_i = \epsilon_i + \mu_k, \text{ for $\tau_{k-1} + 1 \leq i \leq \tau_k$ and $k \in \{1, ..., m+1\}$},
\end{equation}

where $\mu_k \neq \mu_{k+1}$, for $k \in \{1, ..., m\}$ and $(\epsilon_1, ..., \epsilon_n)$ are a set of centered, independent and identically distributed Gaussian random variables. Take a series with $m$ changes and true changepoint locations $\tau_1, ..., \tau_m$ (where $0 < \tau_1 < ... < \tau_m < n$). Apply the PELT procedure, minimising squared error loss, with a penalty of $\beta = \left(2 + \epsilon\right) \log n$, for any $\epsilon > 0$, to produce an estimated set of change locations $0 < \hat{\tau}_1 < ... < \hat{\tau}_{\hat{m}} < n$, some number of estimated changes $\hat{m}$. Then, for any $\alpha > 0$, $\mathbb{P}(\mathcal{E}_n^{\alpha}) \rightarrow 1$ as $n \rightarrow \infty$, where:

\begin{center}
$\mathcal{E}_n^{\alpha} = \left\{\hat{m} = m ; \max\limits_{i = 1, ..., m} |\hat{\tau}_i - \tau_i| \leq \left\lceil \left( \log n \right)^{1 + \alpha}\right\rceil \right\}$.
\end{center}
\end{prop}

\textbf{Proof}: See Section~\ref{sec:appb} of the Supplementary Materials.

This result also extends naturally to the multivariate setting, with a penalty of $\left(d + 1\right) \left(1 + \epsilon\right) \log n$ (see Section~\ref{sec:appb} of the Supplementary Materials for details). For the univariate case, the proof of Proposition~\ref{peltconsistency} follows a similar pattern to that of \cite{Yao88}, though we relax Yao's condition that an upper bound on the estimated number of changes is specified a priori.

\subsection{Consistency and Computational Cost of Chunk and Deal}\label{ssec:compcost}

We now extend the consistency result in the unparallelised setting to obtain equivalent results for Chunk and Deal. These results not only give a bound on the maximum location error of an estimated changepoint, but also provide some insight into the best setting of, for instance, $L(n)$ and $V(n)$, which we can use in turn to provide a theoretical result on the computational power of the new methods. 

\begin{theorem}\label{chunkconsistency}
For the change in mean setting specified in~(\ref{changeinmean}), assume that for a data series of length $n$ we have $L(n)$ cores across which to parallelise a changepoint detection procedure. Defining $\mathcal{E}_n^{\alpha}$ as for the previous results for any $\alpha > 0$, then additionally assuming that $L(n) = o(n)$ with $L(n) \rightarrow \infty$ gives that, under the Chunk procedure for parallelising a procedure which minimises least squared error under a penalty of $\beta = \left(2 + \epsilon\right) \log n$, $\mathbb{P}(\mathcal{E}_n^{\alpha}) \rightarrow 1$ as $n \rightarrow \infty$.
\end{theorem}

\textbf{Proof}: See Appendix.

Note that the definition of the event $\mathcal{E}_n^{\alpha}$ is as for Proposition~\ref{peltconsistency}, meaning that the maximum asymptotic location error under the Chunk procedure is $\mathcal{O}(\log n)$. This error suggests that we can choose $V(n)$ to be small compared to $\frac{n}{L(n)}$, the number of data points on a core. For instance, setting $V(n) = \left\lceil\left(\log n\right)^2 \right\rceil$ ensures accuracy in probability for large $n$ with negligible computational impact relative to $V(n) = \left\lceil\left(\log n\right)\right\rceil$.

In addition, we remark that the conditions on $L(n)$ for Chunk are relatively weak, and are in place to avoid segments of fixed size for $n \rightarrow \infty$. For most practical values of $n$, we advise setting $L(n)$ such that the length of the segments is at least $2 V(n)$ to avoid intersecting overlaps.

\begin{theorem}\label{dealconsistency}
The same result as for Theorem~\ref{chunkconsistency} holds with the Deal parallelisation procedure, assuming in addition that $L(n) \geq \left\lceil\left(\log n\right)^{1 + \alpha}\right\rceil$.
\end{theorem}

\textbf{Proof}: See Appendix.

Note that the conditions on $L(n)$ are stronger for Deal than for Chunk, with a lower bound corresponding with the maximum location error inherent in the event $\mathcal{E}_n^{\alpha}$. We believe the constraint on $L(n)$ is an artefact of the proof technique. Intuitively we would expect the statistical accuracy of Deal to be larger for smaller $L(n)$; as, for example, $L(n) = 1$ corresponds to optimally minimising the cost. Practically, setting $L(n) = \left\lceil\left(\log n\right)\right\rceil$ is unlikely to be problematic for typical values of $n$, a notion which we confirm empirically in Section \ref{sec:sims}.

Finally, given these results, we are now in a position to give a formal statement on the worst case computational cost for both Chunk and Deal, when the computational cost of setting up a parallel environment is assumed to be negligible.

\begin{corollary}\label{parspeed}
Under the change in mean setting outlined in Proposition~\ref{peltconsistency}, with probability tending to 1 as $n \rightarrow \infty$, the computational cost for Chunk when parallelising the PELT procedure using $L(n)$ computer cores is $\mathcal{O}\left(\max\left(\left(\frac{n}{L\left(n\right)}\right)^2, m^2\right)\right)$, while for Deal the cost is \\ $\mathcal{O}\left(\max\left(\left(\frac{n}{L(n)}\right)^2, \left(L(n)\right)^2\right)\right)$, compared to a cost of $\mathcal{O}(n^2)$ for unparallelised PELT.
\end{corollary}

\textbf{Proof}: See Appendix.

We remark that setting $L(n) \sim n^{\frac{1}{2}}$ in Corollary~\ref{parspeed} guarantees a worst case computational cost of $\mathcal{O}(n)$ for both Chunk and Deal, no matter the performance of PELT. In addition, we note that we achieve a computational gain which is quadratic with $L(n)$ in the best case. We emphasise again that this result ignores the cost of setting up a parallel environment, which can lead to PELT performing better computationally for small $n$. Therefore, we now conduct a simulation study in order to understand the likely practical circumstances in which parallelisation is a more efficient option.

\section{Simulations}\label{sec:sims}

We now turn to consider the performance of these parallelised approximate methods on simulated data.

While these suggested parallelisation techniques do speed up the implementation of the dynamic programming procedure underlying, say, PELT, the exactness of PELT in resolving~(\ref{fullcost}) is no longer guaranteed. We therefore compare parallelised PELT with Wild Binary Segmentation (WBS), proposed by \cite{Fryzlewicz14}, a non-exact changepoint method which has impressive computational speed.

\begin{figure}[h!]
\begin{minipage}[t]{\textwidth}
\begin{center}
\includegraphics[width=0.99\textwidth]{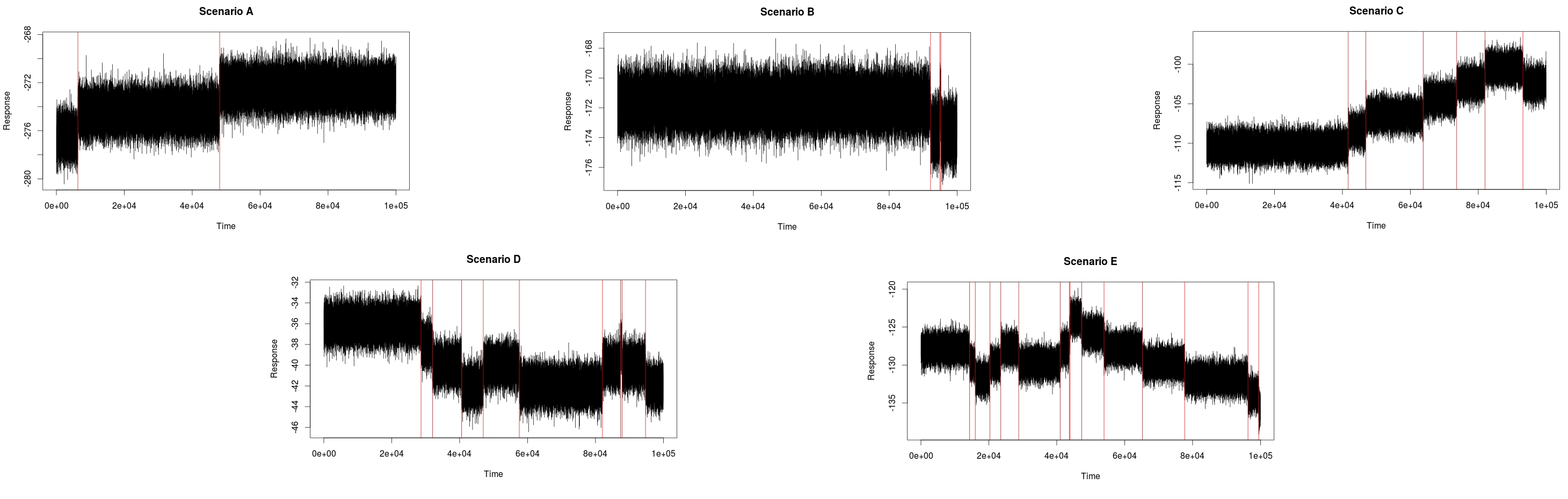}
\end{center}
\vspace{-20pt}
\caption{Five scenarios under examination in the simulation study. Clockwise from top left are scenarios A, B, C, E and D with 2, 3, 6, 14 and 9 changes respectively.}
\label{phase 6}
\end{minipage}
\end{figure}

Simulated time series with piecewise normal segments were generated. Five scenarios, with changes at particular proportions of the time series, were examined in detail in the study. For a time series length of 100000, these scenarios are shown in Figure~\ref{phase 6}.

Different lengths of series for each of the five scenarios, keeping the proportionate change sets the same, were used to examine the statistical power of PELT, Chunk, Deal and WBS under a number of replications for the error terms ($n=200$ in all cases). In addition, four change magnitudes (0.25, 0.5, 1 and 2) were used to examine the behaviour of the algorithms in each of the scenarios as $\Delta \mu$ was increased.

The number of false positives (which were counted as the number of estimated changes more than $\lceil \log n \rceil$ points from the closest true change) and missed changes (the number of true changes with no estimated change within $\lceil \log n \rceil$ points), as well as the maximum observed location error and average location error across all repetitions were measured. Finally, the average cost of the segmentations (using mean squared error) generated by the methods relative to the optimal given by PELT were recorded.

As can be seen from Tables 1 -~\ref{AvLocErr}, Chunk and Deal closely mirror PELT in statistical performance in finding approximately the same number of changes in broadly similar locations. The performance of WBS was generally worse across these measures, although WBS did mirror PELT moderately closely and did occasionally perform better, particularly in certain scenarios for average location error. However, as the number of changes was increased, WBS was generally outperformed by both Chunk and Deal.

\begin{figure}[h!]
\begin{minipage}[t]{\textwidth}
\begin{center}
\includegraphics[width=0.99\textwidth]{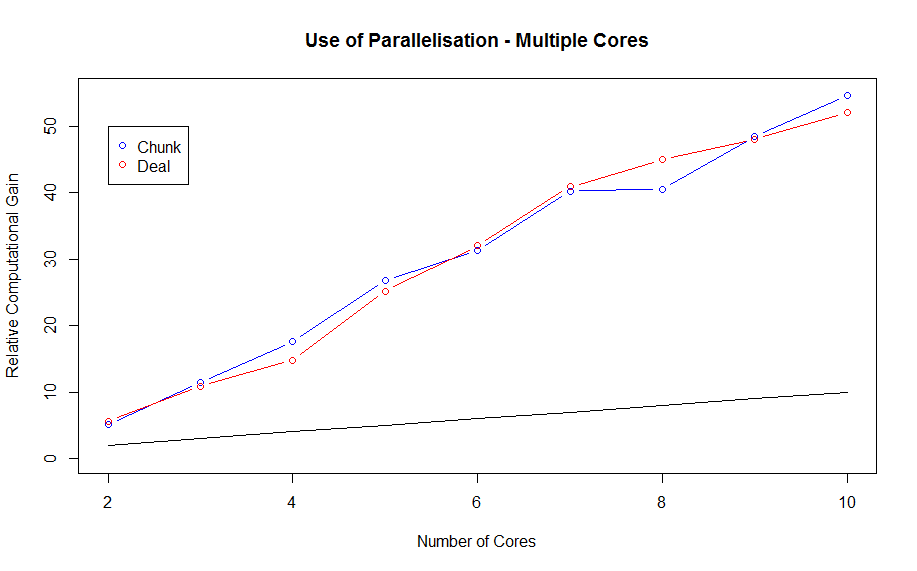}
\end{center}
\vspace{-20pt}
\caption{Relative computational gain for Chunk and Deal across a differing number of cores under Scenario C with $\Delta \mu =1$ and $n=100000$. The line $y=x$ is shown for comparison, demonstrating the super-linear computational gain.}
\label{phase7}
\end{minipage}
\end{figure}

From Table~\ref{AvTimTake}, we note that, in practice, Deal often outperforms Chunk in terms of computational speed for a given number of cores. This is due to the fact that the Deal procedure will rarely perform at the worst case computational speed during the split phase (which typically dominates the computation time), as one of the candidates around a true change is very likely to be chosen as a candidate changepoint (see the proof of Theorem~\ref{dealconsistency}). This means that more candidates for the most recent changepoint are pruned than for Chunk. PELT was observed to be the fastest method for the smallest value of $n$ across all scenarios. It was at the larger values of $n$ where the quadratic gains in speed of Chunk and Deal became apparent, as can also be seen in Figure~\ref{phase7}, which shows the relative computational gain for Scenario C when $\Delta\mu = 1$ and $n=100000$ across multiple different values of $L(n)$.

Finally, from Table~\ref{AvCost}, both Chunk and Deal are seen to track PELT very closely in terms of the final cost of the model. In particular, Deal seems to perform well for smaller values of $\Delta\mu$, while Chunk in general appears to find solutions of a very similar global cost to the PELT algorithm for larger $\Delta \mu$. Caution should be exercised, however, as only the value of $L(n) = 4$ was tested.

\begin{table}
\begin{tabular}{ |p{2.15cm}|p{1.25cm}||p{0.7cm}|p{0.7cm}|p{0.7cm}|p{0.7cm}||p{0.7cm}|p{0.7cm}|p{0.7cm}|p{0.7cm}||p{0.7cm}|p{0.7cm}|p{0.7cm}|p{0.7cm}| }
 \hline
 \multicolumn{2}{|c||}{Average False Alarms} & \multicolumn{4}{|c||}{Length $= 10^3$} & \multicolumn{4}{|c||}{Length $= 10^4$} & \multicolumn{4}{|c|}{Length $= 10^5$} \\
 \hline
  Scenario & Method & 0.25 & 0.5 & 1 & 2 & 0.25 & 0.5 & 1 & 2 & 0.25 & 0.5 & 1 & 2\\
 \hline
A & PELT & 0.69 & 0.77 & \textbf{0.26} & \textbf{0.05} & 1.36 & 0.74 & \textbf{0.15} & 0.01 & 1.28 & \textbf{0.59} & \textbf{0.10} & \textbf{0.00}\\
(2 changes) & Chunk4 & 0.70 & 0.87 & \textbf{0.26} & \textbf{0.05} & 1.50 & 0.74 & 0.16 & 0.01 & 1.29 & \textbf{0.59} & \textbf{0.10} & \textbf{0.00} \\
& Deal4 & 0.68 & 0.73 & \textbf{0.26} & \textbf{0.05} & 1.37 & 0.74 & \textbf{0.15} & 0.01 & 1.37 & 0.74 & 0.15 & 0.01\\
& WBS & \textbf{0.54} & \textbf{0.66} & 0.29 & 0.08 & \textbf{1.20} & \textbf{0.66} & 0.16 & \textbf{0.00} & \textbf{1.26} & \textbf{0.59} & \textbf{0.10} & \textbf{0.00}\\
\hline
B & PELT & 0.17 & 0.29 & \textbf{0.14} & 0.04 & 0.76 & 0.46 & 0.17 & 0.02 & 0.98 & 0.83 & \textbf{0.09} & \textbf{0.00} \\
(3 changes) & Chunk4 & \textbf{0.14} & \textbf{0.23} & 0.17 & 0.04 & 0.72 & 0.46 & 0.15 & \textbf{0.01} & 0.98 & 0.53 & \textbf{0.09} & \textbf{0.00} \\
& Deal4 & 0.16 & 0.27 & 0.16 & \textbf{0.02} & 0.77 & 0.46 & 0.17 & \textbf{0.01} & \textbf{0.77} & \textbf{0.46} & 0.17 & 0.01 \\
& WBS & 0.15 & 0.25 & 0.19 & 0.07 & \textbf{0.55} & \textbf{0.45} & \textbf{0.12} & 0.02 & 0.97 & 0.93 & 0.24 & 0.10 \\
\hline
C & PELT & 0.92 & 1.31 & 0.76 & 0.09 & 3.08 & \textbf{2.10} & \textbf{0.38} & \textbf{0.01} & 3.94 & 1.89 & 0.20 & \textbf{0.00} \\
(6 changes) & Chunk4 & 0.91 & 1.25 & 0.79 & \textbf{0.08} & 2.84 & 2.12 & \textbf{0.38} & \textbf{0.01} & 3.96 & \textbf{1.88} & \textbf{0.19} & \textbf{0.00} \\
& Deal4 & 0.88 & 1.29 & \textbf{0.74} & \textbf{0.08} & 3.07 & 2.17 & 0.39 & \textbf{0.01} & \textbf{3.07} & 2.17 & 0.39 & 0.01\\
& WBS & \textbf{0.86} & \textbf{1.23} & 1.07 & 0.23 & \textbf{2.73} & 2.40 & 0.66 & 0.08 & 4.11 & 2.17 & 0.53 & 0.11\\
\hline
D & PELT & 1.03 & 1.47 & 0.85 & 0.10 & 3.72 & 2.88 & 0.58 & 0.04 & 5.28 & 2.76 & \textbf{0.43} & \textbf{0.01} \\
(9 changes) & Chunk4 & 1.09 & 1.42 & \textbf{0.82} & 0.10 & 3.38 & \textbf{2.85} & \textbf{0.57} & \textbf{0.02} & 5.26 & \textbf{2.75} & \textbf{0.43} & \textbf{0.01} \\
& Deal4 & 1.03 & 1.41 & 0.87 & \textbf{0.08} & 3.73 & 2.89 & 0.60 & 0.03 & \textbf{3.73} & 2.89 & 0.60 & 0.03\\
& WBS & \textbf{0.97} & \textbf{1.27} & 1.01 & 0.17 & \textbf{3.20} & 3.10 & 0.90 & 0.20 & 5.42 & 3.26 & 0.79 & 0.17\\
\hline
E & PELT & 1.04 & 1.72 & \textbf{1.20} & 0.13 & 4.39 & 4.12 & 0.88 & \textbf{0.01} & 8.22 & \textbf{4.12} & \textbf{0.55} & \textbf{0.00}\\
(14 changes) & Chunk4 & 1.09 & \textbf{1.66} & 1.21 & 0.12 & \textbf{4.25} & 4.13 & 0.89 & \textbf{0.01} & 4.38 & \textbf{4.12} & \textbf{0.55} & \textbf{0.00}\\
& Deal4 & 1.04 & \textbf{1.66} & \textbf{1.20} & \textbf{0.10} & 4.32 & \textbf{4.07} & \textbf{0.86} & \textbf{0.01} & \textbf{4.32} & 4.18 & 0.86 & 0.01\\
& WBS & \textbf{1.01} & 1.67 & 1.24 & 0.24 & 3.86 & 4.23 & 1.24 & 0.18 & 8.14 & 4.50 & 1.08 & 0.18\\
 \hline
\end{tabular}
\begin{center}\caption{The average number of false alarms recorded across all 200 repetitions for each of the 5 scenarios A, B, C, D and E. A false alarm is defined as an estimated changepoint which is at least $\left\lceil \left(\log n\right)\right\rceil$ points from the closest true changepoint.}
\end{center}
\label{FalAlarms}
\end{table}

\begin{table}
\begin{tabular}{ |p{2.15cm}|p{1.25cm}||p{0.7cm}|p{0.7cm}|p{0.7cm}|p{0.7cm}||p{0.7cm}|p{0.7cm}|p{0.7cm}|p{0.7cm}||p{0.7cm}|p{0.7cm}|p{0.7cm}|p{0.7cm}| }
 \hline
 \multicolumn{2}{|c||}{Average Num. Missed} & \multicolumn{4}{|c||}{Length $= 10^3$} & \multicolumn{4}{|c||}{Length $= 10^4$} & \multicolumn{4}{|c|}{Length $= 10^5$} \\
 \hline
  Scenario & Method & 0.25 & 0.5 & 1 & 2 & 0.25 & 0.5 & 1 & 2 & 0.25 & 0.5 & 1 & 2\\
 \hline
A & PELT & \textbf{1.77} & 1.10 & 0.22 & \textbf{0.01} & \textbf{1.38} & 0.72 & \textbf{0.14} & \textbf{0.00} & \textbf{1.28} & 0.59 & \textbf{0.10} & \textbf{0.00}\\
(2 changes) & Chunk4 & 1.94 & 1.30 & \textbf{0.21} & \textbf{0.01} & 1.55 & 0.72 & 0.15 & \textbf{0.00} & 1.29 & 0.59 & \textbf{0.10} & \textbf{0.00}\\
& Deal4 & \textbf{1.77} & \textbf{1.09} & 0.22 & \textbf{0.01} & 1.39 & 0.73 & 0.15 & \textbf{0.00} & 1.30 & \textbf{0.58} & \textbf{0.10} & \textbf{0.00} \\
& WBS & 1.84 & 1.29 & 0.22 & \textbf{0.01} & 1.45 & \textbf{0.66} & 0.16 & \textbf{0.00} & 1.26 & 0.59 & \textbf{0.10} & \textbf{0.00}\\
\hline
B & PELT & \textbf{2.62} & \textbf{2.04} & \textbf{1.17} & 1.00 & \textbf{2.47} & \textbf{1.94} & \textbf{1.06} & \textbf{0.00} & 2.45 & \textbf{0.86} & \textbf{0.09} & \textbf{0.00} \\
(3 changes) & Chunk4 & 2.65 & 2.12 & 1.20 & 1.01 & 2.48 & 1.95 & \textbf{1.06} & \textbf{0.00} & 2.45 & \textbf{0.86} & \textbf{0.09} & \textbf{0.00}\\
& Deal4 & 2.65 & 2.08 & 1.19 & 1.02 & 2.50 & \textbf{1.94} & 1.14 & \textbf{0.00} & 2.48 & 0.87 & \textbf{0.09} & \textbf{0.00}\\
& WBS & 2.65 & 2.13 & 1.29 & \textbf{0.91} & 2.51 & 1.95 & \textbf{1.06} & 0.01 & \textbf{2.43} & 1.02 & 0.16 & 0.01\\
\hline
C & PELT & \textbf{5.53} & \textbf{4.55} & 0.74 & \textbf{0.04} & \textbf{4.79} & \textbf{2.08} & \textbf{0.37} & \textbf{0.00} & \textbf{3.94} & 1.89 & 0.20 & \textbf{0.00} \\
(6 changes) & Chunk4 & 5.65 & 4.71 & 0.83 & \textbf{0.04} & 4.91 & 2.10 & \textbf{0.37} & \textbf{0.00} & 3.96 & 1.88 & \textbf{0.19} & \textbf{0.00}\\
& Deal4 & \textbf{5.53} & 4.62 & \textbf{0.73} & \textbf{0.04} & 4.86 & 2.15 & 0.39 & \textbf{0.00} & 3.96 & \textbf{1.01} & \textbf{0.19} & \textbf{0.00} \\
& WBS & 5.57 & 4.71 & 1.22 & 0.08 & 4.90 & 2.36 & 0.56 & 0.03 & 4.05 & 2.08 & 0.48 & 0.04\\
\hline
D & PELT & \textbf{8.20} & \textbf{6.63} & \textbf{2.13} & 0.81 & \textbf{7.45} & 4.32 & 0.72 & 0.02 & 6.38 & 2.75 & \textbf{0.43} & \textbf{0.00}\\
(9 changes)& Chunk4 & 8.36 & 6.74 & 2.20 & 0.83 & 7.62 & \textbf{4.29} & \textbf{0.71} & \textbf{0.01} & \textbf{6.37} & 2.75 & \textbf{0.43} & \textbf{0.00}\\
& Deal4 & 8.22 & 6.68 & 2.23 & 0.84 & 7.50 & 4.34 & 0.77 & 0.02 & 6.39 & \textbf{2.73} & 0.45 & \textbf{0.00}\\
& WBS & 8.22 & 6.66 & 2.65 & \textbf{0.66} & 7.79 & 4.57 & 1.07 & 0.07 & 6.48 & 3.21 & 0.67 & 0.02\\
\hline
E & PELT & \textbf{13.0} & \textbf{11.0} & \textbf{5.18} & \textbf{1.20} & \textbf{12.0} & 6.70 & \textbf{2.07} & \textbf{0.00} & \textbf{9.85} & 4.26 & \textbf{0.55} & \textbf{0.00}\\
(14 changes) & Chunk4 & 13.2 & 11.2 & 5.25 & 1.24 & 12.2 & 6.79 & 2.10 & \textbf{0.00} & 9.90 & \textbf{4.25} & \textbf{0.55} & \textbf{0.00}\\
& Deal4 & \textbf{13.0} & 11.1 & 5.40 & 1.29 & \textbf{12.0} & \textbf{6.68} & 2.09 & \textbf{0.00} & 9.88 & 4.33 & 0.58 & \textbf{0.00}\\
& WBS & 13.1 & 11.2 & 6.09 & 1.53 & 12.3 & 7.46 & 2.51 & 0.16 & 10.2 & 5.00 & 0.97 & 0.04\\
 \hline
\end{tabular}
\caption{The average number of missed changes across all 200 repetitions for each of the 5 scenarios A, B, C, D and E. A missed change is defined as a true changepoint for which no estimated change lies within $\left\lceil \left(\log n\right)\right\rceil$ points.}
\label{AvMissChanges}
\end{table}

\begin{table}
\begin{tabular}{ |p{2.15cm}|p{1.25cm}||p{0.7cm}|p{0.7cm}|p{0.7cm}|p{0.7cm}||p{0.7cm}|p{0.7cm}|p{0.7cm}|p{0.7cm}||p{0.7cm}|p{0.7cm}|p{0.7cm}|p{0.7cm}| }
 \hline
 \multicolumn{2}{|c||}{Average Location Error} & \multicolumn{4}{|c||}{Length $= 10^3$} & \multicolumn{4}{|c||}{Length $= 10^4$} & \multicolumn{4}{|c|}{Length $= 10^5$} \\
 \hline
  Scenario & Method & 0.25 & 0.5 & 1 & 2 & 0.25 & 0.5 & 1 & 2 & 0.25 & 0.5 & 1 & 2\\
 \hline
A & PELT & 64.1 & 21.8 & \textbf{7.43} & 5.96 & 70.0 & 24.7 & 16.1 & 14.4 & \textbf{46.0} & \textbf{11.7} & 3.21 & 1.26\\
(2 changes) & Chunk4 & \textbf{48.3} & 21.9 & 7.63 & 5.67 & 89.5 & 12.4 & 3.63 & 1.24 & 47.4 & \textbf{11.7} & 3.21 & 1.26\\
& Deal4 & 59.1 & \textbf{20.7} & 7.59 & \textbf{5.65} & 56.8 & \textbf{12.0} & \textbf{3.33} & \textbf{1.19} & 46.7 & \textbf{11.7} & 3.20 & \textbf{1.22} \\
& WBS & 86.2 & 34.7 & 12.7 & 10.7 & \textbf{52.4} & 12.3 & 3.40 & 1.20 & \textbf{46.0} & 12.1 & \textbf{3.18} & 1.26\\
\hline
B & PELT & 69.7 & 31.2 & 17.8 & 15.7 & 75.9 & 42.8 & 26.4 & 17.1 & 47.5 & 12.1 & \textbf{3.00} & 1.27 \\
(3 changes) & Chunk4 & 76.5 & 37.1 & 18.3 & 15.7 & 72.4 & 41.6 & 25.3 & 16.3 & \textbf{47.0} & \textbf{10.8} & \textbf{3.00} & 1.27 \\
& Deal4 & \textbf{50.4} & \textbf{22.5} & 17.8 & \textbf{10.0} & 74.8 & 41.2 & 25.9 & 16.4 & 47.8 & 12.4 & \textbf{3.00} & \textbf{1.26}\\
& WBS & 59.9 & 38.7 & \textbf{17.4} & 13.8 & \textbf{32.2} & \textbf{11.0} & \textbf{3.25} & \textbf{1.52} & 47.4 & 14.4 & 5.82 & 3.07\\
\hline
C & PELT & 29.1 & 17.2 & 5.42 & 3.22 & 71.1 & 16.4  & 7.19 & 5.14 & 50.3 & 12.5 & 3.04 & 1.23 \\
(6 changes) & Chunk4 & 31.3 & 18.5 & 4.94 & 2.61 & \textbf{64.1} & 16.4 & 6.64 & 4.59 & 50.7 & 12.4 & 3.01 & 1.24 \\
& Deal4 & 28.1 & 16.7 & \textbf{4.75} & 2.67 & 69.2 & \textbf{16.1} & 6.78 & 4.69 & \textbf{50.2} & \textbf{11.5} & \textbf{2.96} & \textbf{1.18}\\
& WBS & \textbf{21.8} & \textbf{14.1} & 5.87 & \textbf{2.51} & 65.1 & 17.7 & \textbf{5.79} & \textbf{1.88} & 80.7 & 24.0 & 5.62 & 1.93 \\
\hline
D & PELT & 19.3 & 12.2 & 4.37 & 2.80 & 57.3 & 15.0 & 7.98 & 3.20 & \textbf{85.5} & 11.8 & 3.33 & \textbf{1.26} \\
(9 changes) & Chunk4 & 22.3 & 12.6 & 4.67 & 2.48 & 59.5 & \textbf{14.6} & 7.68 & 2.77 & 86.2 & \textbf{11.7} & \textbf{3.32} & \textbf{1.26}\\
& Deal4 & 19.4 & 12.0 & \textbf{4.08} & \textbf{2.40} & \textbf{55.1} & 14.7 & \textbf{4.95} & 2.80 & 85.8 & 11.8 & 3.42 & 1.28\\
& WBS & \textbf{17.6} & \textbf{10.4} & 4.41 & 4.12 & 58.3 & 20.0 & 5.29 & \textbf{1.76} & 199 & 20.4 & 6.47 & 2.39\\
\hline
E & PELT & 14.1 & 9.72 & \textbf{3.83} & 2.09 & 52.0 & \textbf{12.5} & 4.06 & 1.71 & \textbf{51.1} & \textbf{12.2} & \textbf{3.29} & \textbf{1.27} \\
(14 changes) & Chunk4 & 15.3 & \textbf{9.36} & 3.96 & 1.97 & 63.4 & 13.0 & 4.07 & 1.71 & 53.6 & \textbf{12.2} & \textbf{3.29} & 1.29\\
& Deal4 & 14.2 & 9.88 & 3.87 & \textbf{1.70} & \textbf{51.6} & \textbf{12.5} & \textbf{4.03} & 1.75 & \textbf{51.1} & 12.5 & 3.33 & \textbf{1.27} \\
& WBS & \textbf{13.7} & 9.67 & 4.20 & 2.58 & 56.9 & 17.1 & 9.05 & \textbf{1.64} & 70.6 & 36.3 & 5.18 & 1.90\\
 \hline
\end{tabular}
\caption{The average location error between those true changes which were detected by the algorithms and the corresponding estimated change across all 200 repetitions for each of the 5 scenarios.}
\label{AvLocErr}
\end{table}

\begin{table}
\begin{tabular}{ |p{2.15cm}|p{1.25cm}||p{0.7cm}|p{0.7cm}|p{0.7cm}|p{0.7cm}||p{0.7cm}|p{0.7cm}|p{0.7cm}|p{0.7cm}||p{0.7cm}|p{0.7cm}|p{0.7cm}|p{0.7cm}| }
 \hline
 \multicolumn{2}{|c||}{Mean Computational Gain} & \multicolumn{4}{|c||}{Length $= 10^3$} & \multicolumn{4}{|c||}{Length $= 10^4$} & \multicolumn{4}{|c|}{Length $= 10^5$} \\
 \hline
  Scenario & Method & 0.25 & 0.5 & 1 & 2 & 0.25 & 0.5 & 1 & 2 & 0.25 & 0.5 & 1 & 2\\
 \hline
A & Chunk4 & \textbf{0.05} & \textbf{0.05} & \textbf{0.05} & \textbf{0.05} & 2.47 & 2.66 & 2.82 & 2.84 & 12.2 & 10.0 & 12.9 & 14.9 \\
(2 changes) & Deal4 & \textbf{0.05} & \textbf{0.05} & \textbf{0.05} & \textbf{0.05} & \textbf{2.97} & \textbf{3.09} & \textbf{3.25} & \textbf{3.30} & \textbf{21.4} & \textbf{24.2} & \textbf{21.4} & \textbf{22.7} \\
\hline
B & Chunk4 & \textbf{0.05} & \textbf{0.05} & \textbf{0.05} & \textbf{0.05} & 2.75 & 2.78 & 2.74 & 2.68 & 13.4 & 8.55 & 14.4 & 10.7 \\
(3 changes) & Deal4 & \textbf{0.05} & \textbf{0.05} & \textbf{0.05} & \textbf{0.05} & \textbf{2.98} & \textbf{3.20} & \textbf{3.09} & \textbf{3.11} & \textbf{14.2} & \textbf{14.0} & \textbf{34.8} & \textbf{36.3} \\
\hline
C & Chunk4 & \textbf{0.06} & 0.04 & \textbf{0.05} & \textbf{0.06} & 2.72 & 2.97 & 2.79 & 2.77 & 7.65 & 11.1 & \textbf{17.6} & \textbf{10.7} \\
(6 changes) & Deal4 & \textbf{0.06} & \textbf{0.05} & \textbf{0.05} & 0.05 & \textbf{3.05} & \textbf{3.33} & \textbf{3.32} & \textbf{3.16} & \textbf{26.0} & \textbf{32.0} & 14.8 & \textbf{10.7} \\
\hline
D & Chunk4 & \textbf{0.05} & \textbf{0.06} & 0.05 & \textbf{0.08} & 3.17 & 2.96 & 3.22 & 2.96 & 7.91 & 8.21 & 10.2 & 10.7 \\
(9 changes) & Deal4 & \textbf{0.05} & \textbf{0.06} & \textbf{0.06} & \textbf{0.08} & \textbf{3.59} & \textbf{3.95} & \textbf{3.56} & \textbf{3.87} & \textbf{24.3} & \textbf{31.5} & \textbf{26.8} & \textbf{33.2}\\
\hline
E & Chunk4 & \textbf{0.06} & \textbf{0.05} & 0.06 & \textbf{0.05} & 3.10 & 2.94 & 2.69 & 2.48 & 7.00 & 7.09 & 19.8 & 21.4 \\
(14 changes) & Deal4 & \textbf{0.06} & \textbf{0.05} & \textbf{0.07} & \textbf{0.05} & \textbf{3.75} & \textbf{3.61} & \textbf{3.22} & \textbf{3.22} & \textbf{14.1} & \textbf{17.0} & \textbf{12.6} & \textbf{16.4} \\
 \hline
\end{tabular}
\caption{The average relative computational speed of the Chunk and Deal procedures compared to PELT using 4 cores.}
\label{AvTimTake}
\end{table}

\begin{table}
\begin{tabular}{ |p{2.15cm}|p{1.25cm}||p{0.7cm}|p{0.7cm}|p{0.7cm}|p{0.7cm}||p{0.7cm}|p{0.7cm}|p{0.7cm}|p{0.7cm}||p{0.7cm}|p{0.7cm}|p{0.7cm}|p{0.7cm}| }
 \hline
 \multicolumn{2}{|c||}{Average Cost - Optimal} & \multicolumn{4}{|c||}{Length $= 10^3$} & \multicolumn{4}{|c||}{Length $= 10^4$} & \multicolumn{4}{|c|}{Length $= 10^5$} \\
 \hline
  Scenario & Method & 0.25 & 0.5 & 1 & 2 & 0.25 & 0.5 & 1 & 2 & 0.25 & 0.5 & 1 & 2\\
 \hline
A & Chunk4 & 1.69 & 1.12 & \textbf{0.03} & \textbf{0.02} & 3.07 & \textbf{0.05} & \textbf{0.01} & \textbf{0.00} & 0.03 & \textbf{0.00} & \textbf{0.01} & \textbf{0.00}\\
(2 changes) & Deal4 & \textbf{0.10} & \textbf{0.21} & 0.29 & 0.19 & \textbf{0.10} & 0.18 & 0.34 & 0.41 & 0.07 & 0.14 & 0.24 & 0.19\\
& WBS & 1.70 & 1.76 & 0.63 & 0.65 & 1.63 & 0.09 & 0.08 & 0.06 & \textbf{0.00} & 0.10 & 0.10 & 0.10 \\
\hline
B & Chunk4 & 0.13 & 0.59 & \textbf{0.16} & \textbf{0.10} & 0.14 & \textbf{0.01} & \textbf{0.01} & \textbf{0.01} & \textbf{0.00} & \textbf{0.00} & \textbf{0.00} & \textbf{0.00}\\
(3 changes) & Deal4 & \textbf{0.10} & \textbf{0.23} & 0.53 & 2.35 & \textbf{0.13} & 0.23 & 0.62 & 2.52 & 0.09 & 0.28 & 0.49 & 0.52\\
& WBS & 1.04 & 1.60 & 1.67 & 2.18 & 1.23 & 1.01 & 2.09 & 1.93 & 2.00 & 2.00 & 2.80 & 2.80 \\
\hline
C & Chunk4 & 1.47 & 1.33 & \textbf{0.26} & \textbf{0.16} & 3.39 & \textbf{0.18} & \textbf{0.02} & \textbf{0.00} & \textbf{0.05} & \textbf{0.01} & \textbf{0.00} & \textbf{0.00}\\
(6 changes) & Deal4 & \textbf{0.19} & \textbf{0.49} & 1.40 & 4.73 & \textbf{0.22} & 0.58 & 1.25 & 4.48 & 0.26 & 0.46 & 0.64 & 0.66\\
& WBS & 2.12 & 3.79 & 4.14 & 3.72 & 7.01 & 3.51 & 4.18 & 3.97 & 4.30 & 4.80 & 4.30 & 4.50 \\
\hline
D & Chunk4 & 1.99 & 1.11 & \textbf{0.39} & \textbf{0.29} & 3.68 & \textbf{0.05} & \textbf{0.01} & \textbf{0.01} & \textbf{0.05} & \textbf{0.01} & \textbf{0.00} & \textbf{0.00}\\
(9 changes) & Deal4 & \textbf{0.22} & \textbf{0.54} & 1.32 & 2.98 & \textbf{0.31} & 0.77 & 1.69 & 4.97 & 0.37 & 0.76 & 1.82 & 5.42\\
& WBS & 2.09 & 5.01 & 5.38 & 5.26 & 9.03 & 6.12 & 5.94 & 5.94 & 6.60 & 6.60 & 6.30 & 5.10 \\
\hline
E & Chunk4 & 1.96 & 1.75 & \textbf{0.49} & \textbf{0.29} & 5.93 & \textbf{0.53} & \textbf{0.04} & \textbf{0.01} & \textbf{0.09} & \textbf{0.01} & \textbf{0.00} & \textbf{0.00}\\
(14 changes) & Deal4 & \textbf{0.21} & \textbf{0.72} & 2.08 & 4.79 & \textbf{0.42} & 1.16 & 3.19 & 12.0 & 0.58 & 1.16 & 2.79 & 7.49\\
& WBS & 3.06 & 6.38 & 8.95 & 8.04 & 9.76 & 9.84 & 7.52 & 9.37 & 11.8 & 11.0 & 11.2 & 9.00\\
 \hline
\end{tabular}
\caption{The average cost, calculated using the log likelihood of the segments, resulting from executing each procedure. This is adjusted according to the equivalent cost computed by PELT (which is optimal).}
\label{AvCost}
\end{table}

\section{Discussion}\label{sec:discussion}

We have proposed two new methods for changepoint detection, Chunk and Deal, each based on parallelising an existing method, PELT. These methods represent a substantial computational gain in many cases, particularly for large $n$. In addition, by establishing the asymptotic consistency of PELT, we have been able in turn to show the asymptotic consistency of the Chunk and Deal methods, such that the error inherent to all three is $\mathcal{O}\left(\log n\right)$ in terms of the maximum location error of an estimated change relative to the corresponding true change.

We have demonstrated empirically that an implication of this is that Chunk and Deal, while not inheriting the exactness of PELT, do perform well in finding changes in practice.

\section{Acknowledgments}\label{sec:acks}

Tickle is grateful for the support of the EPSRC (grant number EP/L015692/1). The authors also acknowledge British Telecommunications plc (BT) for financial support, and are grateful to Kjeld Jensen and Dave Yearling  in BT Research \& Innovation for helpful discussions.

\appendix
\setcounter{secnumdepth}{0}
\section{Appendix}\label{sec:app}

The following results will be stated with respect to a general $\alpha > 0$. Theoretically, this means that any $\alpha > 0$ can be used in Algorithm~1 or Algorithm~2, however in the simulation study detailed in Section~\ref{sec:sims}, $\left\lceil(\log n)^2\right\rceil$ was used as the overlap length (for Chunk), while the cutoff value for closeness detailed in the merge phase (Step 3) of both procedures was taken as $\left\lceil\left(\log n\right)\right\rceil$.

\textbf{Proof of Theorem~\ref{chunkconsistency}}: It is necessary to establish that:

\vspace{-10pt}

\begin{enumerate}[(I):]
\item Each change is detected by the core in which it is present.
\item At the merge phase, only one estimated change per core is kept.
\end{enumerate}

\textbf{Proof of (I)}: Taking $L(n) = o\left(n\right)$, then $\left\lceil\frac{n}{L(n)}\right\rceil + V(n) \rightarrow \infty$ as $n \rightarrow \infty$. Thus, the Chunk procedure will inherit the asymptotic consistency of the base procedure providing no change consistently falls arbitrarily close to the boundary between two cores. In which case, for $V(n) = \left\lceil\left(\log n\right)^{1 + \alpha}\right\rceil$, the segment length would be reduced from $\mathcal{O}(n)$ to $\mathcal{O}\left(\left\lceil(\log n)^{1 + \alpha}\right\rceil\right)$ (although the smallest possible segment length is $\left\lceil \left(\log n\right)^{1 + \alpha} \right\rceil$). As such a segment length violates the condition on fixed values for $\theta_i = \left\lfloor\tau_i n\right\rfloor$ outlined in Section~\ref{sec:theory}, it is therefore necessary to establish that a true change positioned at a point within $\left\lceil\left(\log n\right)^{1+\alpha}\right\rceil$ of either the beginning or the end of the series will be detected with probability 1 as $n \rightarrow \infty$. It is sufficient to extend Corollary~\ref{errorval} (see Section~\ref{sec:appa} of the Supplementary Materials) to the case where the first true change is at location $\left\lceil \left(\log n\right)^{1 + \alpha} \right\rceil$.

However, as the minimum segment length is at least $\mathcal{O}(n)$, then by the argument of the proof of Corollary~\ref{errorval} each changepoint is detected by at least one core to which it is given in probability for increasing $n$. Formally, one can consider the difference:

\vspace{-30pt}

\begin{align*}
\mbox{Diff} :&= \mbox{RSS}(y_{1:n} ; \hat{\tau}_{1:\hat{m}}) - \mbox{RSS}\left(y_{1:n}; \tau_1 + \left\lceil \left(\log n\right)^{1 + \alpha} \right\rceil, \tau_{2:m}, \hat{\tau}_{1:\hat{m}}\right),
\end{align*}

\vspace{-15pt}

and it can be shown that $\mbox{Diff} = \mathcal{O}((\log n)^{1 + \alpha})$. In particular, $\mbox{Diff} > 2 m\log n$, so the Chunk procedure will detect a change at the boundary with a segment length of $\left \lceil \left(\log n\right)^{1 + \alpha} \right\rceil$.

\textbf{Proof of (II)}: We examine all segmentations, $\hat{\tau}_{1:m}$, such that $\left|\hat{\tau}_i - \tau_i \right| \leq K \log \log n$, for fixed $K > 0$, $\forall i$. Then if $\Delta \mu_k := \left|\mu_k - \mu_{k+1}\right|$:

\vspace{-20pt}

\begin{align*}
\mbox{RSS}(y_{1:n}; \hat{\tau}_{1:m}) - \mbox{RSS}(y_{1:n}; \tau_{1:m}) &\leq \sum_{i = 1}^{m+1} \left\{\frac{1}{\tau_{i} - \tau_{i-1}} \left(\sum_{j = \tau_{i-1} + 1}^{\tau_i} Z_j\right)^2 - \frac{1}{\hat{\tau}_i - \hat{\tau}_{i-1}} \left(\sum_{j = \hat{\tau}_{i-1} + 1}^{\hat{\tau}_i} Z_j\right)^2 \right\} \\
\hspace{7pt}&+ K\sum_{k=1}^m \left(\Delta \mu_k\right)^2 \log \log n + G,
\end{align*}

\vspace{-5pt}

where $G \sim N\left(0,4\sigma^2  K\sum_{k=1}^m \left(\Delta \mu_k\right)^2 \log \log n\right)$.

With recourse once more to the result of \cite{LaurentMassart00} (see equation (\ref{chi}) in Section~\ref{sec:appa} of the Supplementary Materials for the general result), if $U \sim \chi^2_{m+1}$ then:

\begin{align*}
\mathbb{P}(U \geq d \log \log n) &\leq \exp\left(-\frac{d}{2} \log \log n + \sqrt{\left(2d \log \log n - \left(m + 1\right)\right) \left(m + 1\right)}/2\right) \\
&\leq \left(\log n\right)^{-\frac{d}{2} + \delta}, 
\end{align*}

for any $\delta > 0$, providing that $\log \log n > \frac{m + 1}{2d}$. As there are $\left(2K\log \log n\right)^m$ possibilities for the segmentation $\hat{\tau}_{1:m}$, then uniformly across all segmentations, taking $d = 2m$ for example, gives that the difference in the residual sum of squares is uniformly $o_p\left(\log n\right)$.

Now from the proof of Proposition~\ref{peltconsistency} (see Section~\ref{sec:appb} of the Supplementary Materials) we know that in probability any segmentation with more than $m$ changes will have a greater cost than the true segmentation if a penalty of $2\left(1 + \epsilon/2\right)\log n$ is used. Therefore, across all segmentations under consideration here, the cost is at least $\epsilon/2\log n$ greater than the cost of the true segmentation if a penalty of $2\left(1 + \epsilon\right)\log n$ is used. \QEDB

\textbf{Proof of Theorem~\ref{dealconsistency}}: Recall that $L(n) \geq \left\lceil\left(\log n\right)^{1 + \alpha} \right\rceil$ and $L(n) = o(n)$. The idea will be to show that the core which is `dealt' a particular true change, $\tau_i$, will always return this true change as a candidate changepoint for the merge phase. By \cite{Yao88}, letting $\hat{\tau}_{1:m}$ be a set of estimated changes which miss the true change $\tau_i$ by at least $\left\lceil\left(\log n\right)^{1+\alpha}\right\rceil$, then again by the proof of Corollary~\ref{errorval} the cost of this segmentation is strictly worse than the cost of also fitting changes at the points $\tau_i - L(n)$ and $\tau_i + L(n)$. By then considering the difference:

\vspace{-20pt}

\begin{align*}
\mbox{Diff} := \mbox{RSS}(y_{1:n}; \hat{\tau}_{1:m}, \tau_i - L(n), \tau_i + L(n)) - \mbox{RSS}(y_{1:n}; \hat{\tau}_{1:m}, \tau_i - L(n), \tau_i, \tau_i + L(n)),
\end{align*}

in a similar fashion to the proof of Corollary~\ref{errorval}, it can be shown that in probability:

\vspace{-20pt}

\begin{align*}
\frac{\mbox{Diff}}{L(n)} \rightarrow \left(\Delta \mu_{i-1}\right)^2,
\end{align*}

\vspace{-10pt}

where again $\Delta \mu_{i-1}$ is the absolute change in mean at the changepoint $\tau_i$.\QEDB

\textbf{Proof of Corollary~\ref{parspeed}}: It is sufficient to prove the following Claim regarding the number of candidate changes each core returns. 

\textbf{Claim}: In probability, and for any candidate set given to the cores in accordance with the conditions of Theorem~\ref{chunkconsistency} and Theorem~\ref{dealconsistency}:

\begin{enumerate}[(I):]
\item under the Chunk procedure, the maximum number of points returned for the merge phase is bounded above by $2m$,
\item under Deal, the maximum number of points recorded as estimated changes is bounded above by $2m$ for each core.
\end{enumerate}

\textbf{Proof of Claim}:

\textbf{Proof of (I)}: We note that when $L(n)$ is constant, the result is immediate from the proof of Lemma~\ref{peltconsistency}. 

When $L(n) \rightarrow \infty$, it suffices to show that across all cores which are given no true changes, the probability of any of these cores returning a true change converges to 0. Given that the number of cores which are given a change is fixed (and bounded above at $2m$ - as each change could fall inside an overlap), the result is then immediate from the proof of Theorem~\ref{chunkconsistency}.

Considering a single core with no true changes, we adapt the argument from the proof Proposition~\ref{peltconsistency}. For a quantity $U_{k+1}$ which is distributed according to a $\chi^2_{k+1}$ distribution, then by \cite{LaurentMassart00}:

\vspace{-30pt}

\begin{align*}
\mathbb{P}(U_{k+1} \geq d \log n) \leq n^{-\frac{d}{2} + \delta}, \text{ for any } \delta > 0.
\end{align*}

\vspace{-15pt}

Fitting $k > 0$ changes across a core will give that the residual sum of squares relative to a fit of no changes across the same core follows a $\chi^2_{k+1}$ distribution. Therefore, following the application of a Bonferroni correction across all possible placings of $k$ changes gives that the difference between the null fit and the best possible fit of $k$ changes is then bounded in probability as:

\vspace{-20pt}

\begin{align*}
\mathbb{P}(\mbox{Diff} \geq d \log n) \leq n^{-\frac{d}{2} + \delta} \times \left(\frac{n}{L(n)}\right)^k.
\end{align*}

In particular, setting $d = 2\left(1 + \epsilon\right)$ and $\delta = \epsilon/2$ as before, gives that:

\vspace{-10pt}

\begin{align*}
\sum_{k=1}^{n/L\left(n\right)} \mathbb{P}(U_k \geq 2k \left(1 + \epsilon\right) \log n) &\leq \sum_{k=1}^{n/L\left(n\right)} \frac{n^{-\frac{\left(2k - 1\right)\epsilon}{2}}}{\left(L(n)\right)^k} \\
&= \frac{n^{-\frac{\epsilon}{2}}}{L(n)}\left(\frac{1 - n^{-\epsilon\frac{n}{L\left(n\right)}}L(n)^{-\frac{n}{L\left(n\right)}}}{1 - n^{-\epsilon}L(n)^{-1}}\right) \rightarrow 0, \text{ } \forall \epsilon > 0,
\end{align*}

and so scaling this by $L(n)$:

\vspace{-25pt}

\begin{align*}
\mathbb{P}(\text{A core with no true changes overfits}) &\rightarrow 0 \text{ } \forall \epsilon > 0.
\end{align*}

\vspace{-10pt}

Therefore, the computation time of the merge phase of Chunk is $\mathcal{O}(m^2)$ in the worst case, which along with the worst case cost from the split phase of $\mathcal{O}\left(\left(\frac{n}{L\left(n\right)}\right)^2\right)$ gives the worst case computation time for the whole procedure.

\textbf{Proof of (II)}: Define, for a given core under the Deal procedure:

\vspace{-25pt}

\begin{align*}
\mathcal{S}_2 = \left\{s_1^{\left(1\right)}, s_1^{\left(2\right)}, s_2^{\left(1\right)}, s_2^{\left(2\right)}, \ldots, s_m^{\left(1\right)}, s_m^{\left(2\right)}\right\},
\end{align*}

\vspace{-10pt}

where $s_i^{\left(1\right)}$ is the final point given to the core which is strictly before $\tau_i$, and $s_i^{\left(2\right)}$ is the first point given to the core which is after $\tau_i$. In the same way as for the proof of Proposition~\ref{peltconsistency}, we examine the best possible segmentations which include $\mathcal{S}_2$ as a subset of the estimated changepoints for a core, and show that all are rejected in favour of $\mathcal{S}_2$ in probability. We then show that this is true across all cores in probability. 

For a given core, suppose $\mathcal{S}_3$ is a set of points estimated as changes under the Deal procedure such that $\mathcal{S}_2 \subset \mathcal{S}_3$. By construction of $\mathcal{S}_2$, all points in $\mathcal{S}_3 \cap \mathcal{S}_2^c$ must lie in a region between two points of $\mathcal{S}_2$ which also does not contain any true changes. We can therefore apply the same argument as for Proposition~\ref{peltconsistency} to the difference:

\vspace{-30pt}

\begin{align*}
\mbox{Diff} := \mbox{RSS}(y_{\mathcal{A}}; \mathcal{S}_2) - \mbox{RSS}(y_{\mathcal{A}}; \mathcal{S}_3),
\end{align*}

\vspace{-15pt}

where $\mathcal{A}$ refers to any such region between two consecutive points of $\mathcal{S}_2$ which contains a point found only in $\mathcal{S}_3$. Uniformly across such regions, and supposing $k > 0$ such estimated changes are found within $\mathcal{A}$, it can be seen that the positive term in the expression of the difference above is distributed as $\chi^2_{k+1}$. Thus letting $\tilde{n} = \frac{n}{L\left(n\right)}$ and again with recourse to the Bonferroni correction argument as in Proposition~\ref{peltconsistency}, for a given $\epsilon > 0$:

\vspace{-25pt}

\begin{align*}
\sum_{k=1}^{\tilde{n}} \mathbb{P}(\mbox{Diff} \geq 2k\left(1 + \epsilon\right) \log n) &\leq \sum_{k=1}^{\tilde{n}} \frac{n^{-\frac{\left(2k - 1\right)\epsilon}{2}}}{\left(L(n)\right)^k} \\
&= \frac{n^{-\frac{\epsilon}{2}}}{L(n)} \left(\frac{1 - n^{-\tilde{n}\epsilon}L(n)^{-\tilde{n}}}{1 - n^{-\epsilon} L(n)^{-1}}\right) \rightarrow 0, \text{ } \forall \epsilon > 0.
\end{align*}

\vspace{-10pt}

Note that this argument does not consider segmentations which do not contain $\mathcal{S}_2$ as a proper subset. In order to extend this argument, we define the following three sets of segmentations (with respect to a given core):

\vspace{-25pt}

\begin{align*}
\mathcal{GS}_2 &= \left\{\bm{\hat{\tau}} : \left|\bm{\hat{\tau}}\right| = 2m;  \hat{\tau}_{2t - 1} \leq \tau_t, \hat{\tau}_{2t} > \tau_t, \forall t \in \left\{1, ..., m \right\}\right\},\\
\mathcal{GS}_1 &= \left\{\bm{\hat{\tau}} : \left|\bm{\hat{\tau}}\right| \leq 2m; \left|\bm{\hat{\tau}} \cap \left\{\tau_t +1, ..., \tau_{t+1} \right\}\right| \geq 1, \forall t \in \left\{0, ..., m \right\}; \left|\bm{\hat{\tau}} \cap \left\{\tau_t +1, ..., \tau_{t+1} \right\}\right| = 1, \text{ some } t \notin \{0, m\}\right\},\\
\mathcal{GS}_0 &= \left\{\bm{\hat{\tau}} : \left|\bm{\hat{\tau}}\right| \leq 2m; \left|\bm{\hat{\tau}} \cap \left\{\tau_t +1, ..., \tau_{t+1} \right\}\right| = 0, \text{ some } t \right\}.
\end{align*}

Note that $\mathcal{S}_2 \in \mathcal{GS}_2$ and that the argument showing that any segmentation $\mathcal{S}_3$ containing $\mathcal{S}_2$ is rejected uniformly in favour of $\mathcal{S}_2$ may be extended to any element of $\mathcal{GS}_2$ to show that any segmentation with more than $2m$ estimated changes in total and which has at least two estimated changes between each true change is uniformly dominated by a corresponding element of $\mathcal{GS}_2$. 

In the same way, let us now consider extensions from a general element, $\mathcal{T}_1 \in \mathcal{GS}_1$, where here an extension is defined as a superset of $\mathcal{T}_1$ which also contains additional estimated changes from regions between two estimated changes within $\mathcal{T}_1$ not containing a true change. Letting, for example:

\vspace{-25pt}

\begin{align*}
\mathcal{T}_1 = \left\{s_1^{\left(1\right)}, s_1^{\left(2\right)}, ..., s_{i-1}^{\left(2\right)}, s_i^{\left(k\right)}, s_{i+1}^{\left(1\right)}, ..., s_m^{\left(2\right)} \right\} \subset \mathcal{S}_2,
\end{align*}

for some $k \in \left\{1, 2\right\}$ and $i \in \left\{1, ..., m\right\}$. Then any extensions of $\mathcal{T}_1$ consists of placing any further estimated changes in any of the regions between the changes above with the exception of either (if $k =1$) the region $\left(s_i^{\left(1\right)}, s_{i+1}^{\left(1\right)}\right)$ or (if $k=2$) the region $\left(s_{i-1}^{\left(2\right)}, s_{i}^{\left(2\right)}\right)$. Let $\mathcal{T}_1'$ be an arbitrary such extension, and again let $\mathcal{A}$ be any region between two consecutive points of $\mathcal{T}_1$ which contains a point found only in $\mathcal{T}_1'$. As before, uniformly across such regions, and supposing again that $k>0$ such estimated changes are found within $\mathcal{A}$, letting:

\vspace{-30pt}

\begin{align*}
\mbox{Diff} := \mbox{RSS}(y_{\mathcal{A}};\mathcal{T}_1) - \mbox{RSS}(y_{\mathcal{A}}; \mathcal{T}_1'),
\end{align*}

\vspace{-10pt}

then again Diff is distributed as $\chi^2_{k+1}$. With recourse to the same argument as before (noting again that any such region $\mathcal{A}$ will have at most $\tilde{n} = \frac{n}{L\left(n\right)}$ candidate points for the extension - no matter which base element of $\mathcal{GS}_1$ we pick), and extending to other elements of $\mathcal{GS}_1$, we conclude that any segmentation with more than $2m$ estimated changes which places just one estimated change between two true changes in at least one case will be rejected uniformly (and for all cores) in favour of an element of $\mathcal{GS}_1$.

Finally, we consider all segmentations with more than $2m$ changes which place no estimated changes between two true changes in at least one case. We again compare with $\mathcal{T}_0 \in \mathcal{GS}_0$. Letting, for example:

\vspace{-25pt}

\begin{align*}
\mathcal{T}_0 = \left\{s_1^{\left(1\right)}, s_2^{\left(2\right)}, \ldots, s_{i-1}^{\left(2\right)}, s_{i+1}^{\left(1\right)}, \ldots, s_m^{\left(2\right)} \right\},
\end{align*}

for some $i \in \left\{1, \ldots, m \right\}$. Then any extensions of $\mathcal{T}_0$ consists of placing any further estimated changes in any of the regions between the changes above with the exception of the region $\left(s_{i-1}^{\left(2\right)}, s_{i+1}^{\left(1\right)} \right)$. Let $\mathcal{T}_0'$ be an arbitrary such extension, and again let $\mathcal{A}$ be any region between two consecutive points of $\mathcal{T}_0$ which contains a point found only in $\mathcal{T}_0'$. Then again letting:

\vspace{-30pt}

\begin{align*}
\mbox{Diff} := \mbox{RSS}(y_\mathcal{A}; \mathcal{T}_0) - \mbox{RSS}(y_\mathcal{A}; \mathcal{T}_0'),
\end{align*}

\vspace{-10pt}

then for $k>0$ changes in the region $\mathcal{A}$, Diff is distributed as $\chi^2_{k+1}$. We can again extend this argument to extensions of other elements of $\mathcal{GS}_0$ to conclude that segmentations with more than $2m$ changes which have no estimated changepoints between two consecutive true changes in at least one case will be uniformly rejected in favour of an element of $\mathcal{GS}_0$.

Therefore, as any segmentation with more than $2m$ changes for any core is an extension of an element of $\mathcal{GS}_0$, $\mathcal{GS}_1$ or $\mathcal{GS}_2$ (as such a segmentation must contain a region between two consecutive true changes with at least three estimated changes), then across all cores, a segmentation must be picked from within one of the classes $\mathcal{GS}_0$, $\mathcal{GS}_1$ or $\mathcal{GS}_2$ in probability. Thus, the maximum number of estimated changepoints that a core can return in the Deal procedure is $2m$.

The number of candidates returned for the merge phase of the Deal procedure is therefore bounded in probability by $2m L(n)$, so that the maximum computation time of the merge phase is $\mathcal{O}\left(\left(L(n)\right)^2\right)$ in the worst case, giving the total worst case computation time for the whole procedure. \QEDB

\bibliography{bibliography}

\begin{thebibliography}{}

\bibitem[\protect\astroncite{Alba}{2005}]{Alba05}
Alba, E. (2005).
\newblock {\em Parallel Metaheuristics}.
\newblock John Wiley \& Sons, Inc., Hoboken, New Jersey, United States of
  America.

\bibitem[\protect\astroncite{Chen and Nkurunziza}{2017}]{ChenNkurunziza17}
Chen, F. and Nkurunziza, S. (2017).
\newblock On estimation of the change points in multivariate regression models
  with structural changes.
\newblock {\em Communications in Statistics - Theory and Methods}, 46(14):7157
  -- 7173.

\bibitem[\protect\astroncite{Chen and Gupta}{2000}]{ChenGupta00}
Chen, J. and Gupta, A.~K. (2000).
\newblock {\em Parametric Statistical Changepoint Analysis}.
\newblock Birkhäuser, Boston, Massachusetts, United States of America.

\bibitem[\protect\astroncite{Fearnhead and Rigaill}{2017}]{FearnheadRigaill17}
Fearnhead, P. and Rigaill, G. (2017).
\newblock Changepoint detection in the presence of outliers.
\newblock {\em arXiv:1609.07363v2}, pages 1 -- 29.

\bibitem[\protect\astroncite{Fryzlewicz}{2014}]{Fryzlewicz14}
Fryzlewicz, P. (2014).
\newblock Wild binary segmentation for multiple change-point detection.
\newblock {\em The Annals of Statistics}, 42(6):2243--2281.

\bibitem[\protect\astroncite{Haynes et~al.}{2017}]{HaynesEckleyFearnhead17}
Haynes, K., Eckley, I., and Fearnhead, P. (2017).
\newblock Computationally efficient changepoint detection for a range of
  penalties.
\newblock {\em Journal of Computational and Graphical Statistics},
  26(1):134--143.

\bibitem[\protect\astroncite{Jackson et~al.}{2005}]{JacksonScargleBarnes05}
Jackson, B., Scargle, J., Barnes, D., Arabhi, S., Alt, A., Gioumousis, P.,
  Gwin, E., Sangtrakulcharoen, P., Tan, L., and Tsai, T. (2005).
\newblock An algorithm for optimal partitioning of data on an interval.
\newblock {\em IEEE Signal Processing}, 12(2):105--108.

\bibitem[\protect\astroncite{Killick et~al.}{2012}]{KillickFearnheadEckley12}
Killick, R., Fearnhead, P., and Eckley, I. (2012).
\newblock Optimal detection of changepoints with a linear computational cost.
\newblock {\em Journal of the American Statistical Association},
  107(500):1590--1598.

\bibitem[\protect\astroncite{Laurent and Massart}{2000}]{LaurentMassart00}
Laurent, B. and Massart, P. (2000).
\newblock Adaptive estimation of a quadratic functional by model selection.
\newblock {\em The Annals of Statistics}, 28(5):1302 -- 1338.

\bibitem[\protect\astroncite{Mezmaz et~al.}{2011}]{Mezmaz11}
Mezmaz, M., Melab, M., Kessaci, Y., Lee, Y., Talbi, E.-G., Zomaya, A., and
  Tuyttnes, D. (2011).
\newblock A parallel bi-objective hypbrid metaheuristic for energy-aware
  scheduling for cloud computing systems.
\newblock {\em Journal of Parallel and Distributed Computing},
  71(11):1497--1508.

\bibitem[\protect\astroncite{Rigaill et~al.}{2013}]{Hocking13}
Rigaill, G., Hocking, T.~D., Bach, F., and Vert, J.~P. (2013).
\newblock Learning sparse penalties for change-point detection using max margin
  interval regression.
\newblock {\em Proceedings of the 30th International Conference on Machine
  Learning (ICML-13)}.

\bibitem[\protect\astroncite{Rigaill et~al.}{2012}]{Rigaill12}
Rigaill, G., Lebarbier, E., and Robin, S. (2012).
\newblock Exact posterior distributions and model selection criteria for
  multiple change-point detection problems.
\newblock {\em Statistics and Computing}, 22(4):917--929.

\bibitem[\protect\astroncite{Schmid et~al.}{2012}]{Schmid12}
Schmid, N., Christ, C., Christen, M., Eichenberger, A., and van Gunsteren, W.
  (2012).
\newblock Architecture, implementation and parallelisation of the gromos
  software for biomolecular simulation.
\newblock {\em Computer Physics Communications}, 183(4):890--903.

\bibitem[\protect\astroncite{Scott and Knott}{1974}]{ScottKnott74}
Scott, A. and Knott, M. (1974).
\newblock A cluster analysis method for grouping means in the analysis of
  variance.
\newblock {\em Biometrics}, 30(3):507--512.

\bibitem[\protect\astroncite{Truong et~al.}{2017}]{Truong17}
Truong, C., Gudre, L., and Vayatis, N. (2017).
\newblock Penalty learning for changepoint detection.
\newblock {\em Proceedings of the 2017 25th European Signal Processing
  Conference (EUSIPCO) in Kos, Greece}.

\bibitem[\protect\astroncite{Truong et~al.}{2018}]{Truong18}
Truong, C., Oudre, L., and Vayatis, N. (2018).
\newblock A review of changepoint detection methods.
\newblock {\em arXiv:1801.00718}, pages 1--31.

\bibitem[\protect\astroncite{Wang and Dunson}{2014}]{WangDunson14}
Wang, X. and Dunson, D.~B. (2014).
\newblock Parallelizing mcmc via weierstrass sampler.
\newblock {\em arXiv:1312.4605v2}, pages 1--35.

\bibitem[\protect\astroncite{Yao}{1988}]{Yao88}
Yao, Y.-C. (1988).
\newblock Estimating the number of change-points via schwarz' criterion.
\newblock {\em Statistics \& Probability Letters}, 6(3):181--189.

\bibitem[\protect\astroncite{Yao and Au}{1989}]{YaoAu89}
Yao, Y.-C. and Au, S.~T. (1989).
\newblock Least-squares estimation of a step function.
\newblock {\em Sankhya: The Indian Journal of Statistics, Series A},
  51(3):370--381.

\end{thebibliography}

\newpage

\setcounter{secnumdepth}{1}

\begin{center}
\large Parallelisation of a Common Changepoint Detection Method \\
\huge Supplementary Materials \\
\large S. O. Tickle, I. A. Eckley, P. Fearnhead, K. Haynes \\
October 8, 2018
\end{center}

\normalsize

\section{Yao's Results and Extension}\label{sec:appa}

The following two lemmas are due to \cite{Yao88}.

\begin{lemma}
\label{Yao1}
Suppose $Z_1, ..., Z_n \sim^{i.i.d.} N(0, \sigma^2)$. Then for any $\epsilon > 0$ as $n \rightarrow \infty$:

\begin{equation}
\label{Y1}
\mathbb{P}\left(\max_{0 \leq i < j \leq n} \frac{\left(Z_{i+1} + ... + Z_j\right)^2}{\left(j - i\right)} > 2\left(1 + \epsilon\right) \sigma^2 \log n\right) \rightarrow 0.
\end{equation}
\end{lemma}

\begin{lemma}
\label{Yao4}
 Let $m_U$ be an upper bound on the number of changes, and let $\left(\hat{\tau}_1, ..., \hat{\tau}_{\hat{m}}\right)$ be the set of estimated changes generated (by Yao's procedure). For every $\hat{m}$ s.t. $m < \hat{m} \leq m_U$ and $1 \leq r \leq m$,

\begin{center}
$\mathbb{P}( \left( \hat{\tau}_1, ..., \hat{\tau}_{\hat{m}} \right) \in B_i^2\left( n \right)) \rightarrow 0$
\end{center}

as $n \rightarrow \infty$, where:

\begin{center}
$B_i^{\delta}(n) = \{ \left(\xi_1, ..., \xi_t\right): 0 < \xi_1 < ... < \xi_t < n$ and $\left|\xi_s - \tau_r\right| \geq \left\lceil \left(\log n\right)^{\delta} \right\rceil$ for $1 \leq s \leq \hat{m} \}$. 
\end{center}
\end{lemma}

\begin{corollary}
\label{errorval}
Lemma~\ref{Yao4} can be extended to $B_i^{1+\alpha}(n)$, for any $\alpha > 0$.
\end{corollary}

\textbf{Proof of Corollary~\ref{errorval}}: The argument for the location accuracy being $\left(\log n\right)^2$ in \cite{Yao88} comes from showing that the residual sum of squares for a segmentation that misses a change by more than this amount can be reduced by an amount that is greater than $3\left(2 + \epsilon\right)\log n$ with probability tending to $1$ as $n$ increases, by adding three changes at the changepoint plus or minus $\left(\log n\right)^2$. Thus such a segmentation cannot be optimal as the penalised cost for the latter segmentation will be less than the original one. We therefore need only show that this argument holds if we replace an accuracy of $\left(\log n\right)^2$ with $\left(\log n\right)^{1 + \alpha}$ for any $\alpha > 0$.

To do this it suffices to show that a segmentation $\hat{\tau}_1, ..., \hat{\tau}_{\hat{m}}$ which misses a particular change $\tau_i$ by at least $\left\lceil \left( \log n\right)^{1 + \alpha} \right\rceil$ has a residual sum of squares between the points $\tau_i - \left\lceil\left(\log n\right)^{1 + \alpha}\right\rceil$ and $\tau_i + \left\lceil\left(\log n\right)^{1 + \alpha} \right\rceil$ which when normalised by the true fit has term of leading order $\left\lceil \left(\log n\right)^{1 + \alpha} \right\rceil$.

For a segmentation $\hat{\tau}_{1:\hat{m}}$ define $\mbox{RSS}(y_{s:t};\hat{\tau}_{1:\hat{m}})$ to be the residual sum of squares obtained if we fit the changepoints to the subset of data $y_{s:t}$. Note that this will only depend on the changepoints, if any, that lie between timepoints $s$ and $t$. Then for any $\hat{\tau}_{1:\hat{m}} \in B_i^{1+\alpha}(n)$:

\begin{equation}\label{inequalityone}
\mbox{RSS}(y_{1:n};\hat{\tau}_{1:\hat{m}}) \geq \mbox{RSS}\left(y_{1:n};\hat{\tau}_{1:\hat{m}}, \tau_1, \ldots, \tau_{i-1}, \tau_i - \left\lceil\left(\log n\right)^{1+\alpha}\right\rceil, \tau_i + \left\lceil\left(\log n\right)^{1 + \alpha} \right\rceil, \tau_{i+1}, \ldots, \tau_m\right).
\end{equation}

As~\cite{Yao88} remarks, RHS of~(\ref{inequalityone}) can be decomposed as:

$\mbox{RSS}(y_{1:\tau_1}; \mathcal{T}_1) + \ldots + \mbox{RSS}\left(y_{\tau_{i-1}+1:\tau_i - \left\lceil\left(\log n\right)^{1 + \alpha}\right\rceil}; \mathcal{T}_i\right) + \mbox{RSS}\left(y_{\tau_i - \left\lceil\left(\log n\right)^{1 + \alpha}\right\rceil+1:\tau_i + \left\lceil\left(\log n\right)^{1 + \alpha} \right\rceil}; \emptyset\right)$

$+ \mbox{RSS}\left(y_{\tau_i+\left\lceil\left(\log n\right)^{1 + \alpha} \right\rceil+1:\tau_{i+1}}; \mathcal{T}_{i+1}\right) + \ldots + \mbox{RSS}\left(y_{\tau_m+1:n};\mathcal{T}_{m+1}\right)$,

where $\mathcal{T}_h$ is the subset of $\hat{\tau}_{1:\hat{m}}$ which falls inside the corresponding segment of the univariate time series. By Lemma~\ref{Yao1}, each term in this decomposition involving $\mathcal{T}_h$ is such that:

\begin{align*}
\mbox{RSS}(y_{a+1:b};\mathcal{T}_h) = \sum_{j=a+1}^b Z_j^2 + O_p(\log n),
\end{align*}

while, if without loss of generality we assume that the mean at the changepoint $\tau_i$ changes from $0$ to $\mu$, then letting $c_n^{\left(\alpha\right)}= \left\lceil \left(\log n\right)^{1 + \alpha} \right\rceil$:

\begin{align*}
\mbox{RSS}\left(y_{\tau_i - c_n^{\left(\alpha\right)}+1:\tau_i + c_n^{\left(\alpha\right)}}; \emptyset\right) &= \sum_{j=\tau_i - c_n^{\left(\alpha\right)}+1}^{\tau_i + c_n^{\left(\alpha\right)}} \left(Y_i - \bar{Y}\left(\tau_i - c_n^{\left(\alpha\right)}+1,\tau_i + c_n^{\left(\alpha\right)}\right)\right)^2 \\
&=\sum_{j=\tau_i - c_n^{\left(\alpha\right)}+1}^{\tau_i + c_n^{\left(\alpha\right)}} Z_j^2 + \frac{\mu^2}{2}c_n^{\left(\alpha\right)} - \frac{1}{2c_n^{\left(\alpha\right)}} \left(\sum_{j = \tau_i - c_n^{\left(\alpha\right)} + 1}^{\tau_i + c_n^{\left(\alpha\right)}} Z_j\right)^2 + D,
\end{align*}

where $D \sim N\left(0, 2\sigma^2c_n^{\left(\alpha\right)} \mu^2\right)$. Therefore:

\begin{align*}
\left\{\sum_{j = \tau_i - c_n^{\left(\alpha\right)} + 1}^{\tau_i + c_n^{\left(\alpha\right)}} Z_j^2 - \mbox{RSS}\left(y_{\tau_i - c_n^{\left(\alpha\right)}+1:\tau_i + c_n^{\left(\alpha\right)}}; \emptyset\right)\right\}/c_n^{\left(\alpha\right)} &= \frac{\mu^2}{2} - \frac{1}{2\left(c_n^{\left(\alpha\right)}\right)^2}\left(\sum_{j = \tau_i - c_n^{\left(\alpha\right)} + 1}^{\tau_i + c_n^{\left(\alpha\right)}} Z_j\right)^2 + D/c_n^{\left(\alpha\right)} \\
&\rightarrow \frac{\mu^2}{2} \text{ by Lemma~\ref{Yao1}.}
\end{align*}

In particular, $\forall \hat{\tau}_{1:\hat{m}} \in B_i^{1+\alpha}(n)$:

\begin{align*}
\left\{\mbox{RSS}\left(x_{1:n};\hat{\tau}_{1:\hat{m}},\tau_{1:n}^{-i}, \tau_i - c_n^{\left(\alpha\right)}, \tau_i + c_n^{\left(\alpha\right)}\right) - \sum_{j=1}^n Z_j^2 \right\}/c_n^{\left(\alpha\right)} \rightarrow \frac{\mu^2}{2}.
\end{align*}

Thus, as any segmentation from $B_i^{1+\alpha}(n)$ is strictly worse than a corresponding segmentation, which in turn is worse (in probability) than fitting the truth under a penalty of $\beta = 2\left(1+\epsilon\right) \log n$, uniformly in $B_i^{1+\alpha}(n)$, $\mathbb{P}(\hat{\tau}_{1:\hat{m}} \in B_i^{1+\alpha}(n)) \rightarrow 0$. \QEDB

\section{Unparallelised Consistency Results}\label{sec:appb}

\textbf{Proof of Proposition~\ref{peltconsistency}}: Let $\hat{m}$ be the number of changes estimated by the procedure. The aim is firstly to show that:

\textbf{(a)}: $\mathbb{P}(\hat{m} > m) \rightarrow 0$,

\textbf{(b)}: $\mathbb{P}(\hat{m} < m) \rightarrow 0$.

\textbf{Proof of (a)}: Under Corollary~\ref{errorval}, for $\hat{m} > m$, with probability $1$ as $n \rightarrow \infty$ it must be the case that $m$ of the estimated changes are within $\left(\log n\right)^{1 + \alpha}$, some $\alpha > 0$, of the true changes. We will now show that with probability tending to $1$ these segmentations cannot be optimal.

To do this we will compare the penalised cost of any such segmentation with the penalised cost of the true segmentation. The latter cost can be bounded above by $\sum\limits_{t=1}^n Z_t^2 + m\left(2 + \epsilon\right)\log n$. Our approach is to split the comparison of the residual sum of squares of a segmentation $\hat{\tau}_{1:\hat{m}}$ with $\sum\limits_{t=1}^n Z_t^2$ into comparisons for a fixed number of regions of data. To do this, define $c_n^{\left(\alpha\right)}=\left\lceil \left(\log n\right)^{1 + \alpha} \right\rceil$, $u_0=0$, $l_{m+1}=n$, and for $i=1,\ldots,m$, $l_i=\tau_i-c_n^{\left(\alpha\right)}$ and $u_i=\tau_i+c_n^{\left(\alpha\right)}$. We can partition the time points $1, \ldots, n$ into regions $\mathcal{B}_i = \{u_{i-1}+1, \ldots, l_i \}$, for $i=1, \ldots, m+1$ and regions $\mathcal{L}_i = \{l_i+1, \ldots, \tau_i\}$ and $\mathcal{R}_i = \{\tau_i+1,\ldots, u_i\}$ for $i=1, \ldots, m$. These can be viewed as regions more than $c_n^{\left(\alpha\right)}$ from a changepoint, and regions of length $c_n^{\left(\alpha\right)}$ that are respectively left and right of a changepoint.

It is straightforward to show that for any segmentation:

\[
\mbox{RSS}(y_{1:n};\hat{\tau}_{1:\hat{m}})\geq \sum_{i=1}^{m+1} \mbox{RSS}(y_{\mathcal{B}_i};\hat{\tau}_{1:\hat{m}})
+\sum_{i=1}^{m+1} \mbox{RSS}(y_{\mathcal{L}_i};\hat{\tau}_{1:\hat{m}})
+\sum_{i=1}^{m+1} \mbox{RSS}(y_{\mathcal{R}_i};\hat{\tau}_{1:\hat{m}}).
\]

The proof proceeds by showing that on each region $\mathcal{B}_i$ if we have $k = k(\hat{\tau}_{1:\hat{m}})$ changepoints that lie within this region then with probability tending to 1:

\[
\max_{\hat{\tau}_{1:\hat{m}}} \left\{\mbox{RSS}(y_{\mathcal{B}_i};\hat{\tau}_{1:\hat{m}}) + 2\left(1+\epsilon/2\right)k\log n - \sum_{t=u_i+1}^{l_i} Z_t^2\right\} > -4\log\log n.
\]

Then we show that on each region $\mathcal{L}_i$ (and similarly each region $\mathcal{R}_i$) that if there are $k = k(\hat{\tau}_{1:\hat{m}})$ changepoints, then with probability tending to 1:

\[
\max_{\hat{\tau}_{1:\hat{m}}} \left\{ \mbox{RSS}(y_{\mathcal{L}_i};\hat{\tau}_{1:\hat{m}}) - \sum_{t=u_i+1}^{l_i} Z_t^2 \right\} > - 4 \left(k + 1\right) \log\log n.
\]

Taken together we have, with probability tending to 1, a uniform bound on the difference in cost between any segmentation with more than $m$ changepoints, that has one change within $c_n^{\left(\alpha\right)}$ of each true change, and the true segmentation. As such a segmentation can only have, at most, $\hat{m} - m$ changes in regions $\mathcal{B}_i$ this difference is bounded by:

\[
(\hat{m}-m)\epsilon \log n- 4(2m+3)\log \log n>\epsilon \log n - 4(2m+3)\log \log n,
\]
which is positive for large enough $n$.

Note that on each region $\mathcal{B}_i, \mathcal{L}_i, \mathcal{R}_i$ there are no true changes so any estimated changes we do fit inside these regions will involve fitting changes to the noise. Take a generic region of length $\tilde{n}$ which contains no true changes. We examine the reduction in the residual sum of squares when we add $0$ and $k>0$ estimated changes. Note that in the former case it is true that:

\[
- \mbox{RSS}(y_{\mathcal{A}_i}; \hat{\tau}_{1:\hat{m}}) + \sum_{t = a_i + 1}^{a_i + \tilde{n}} Z_t^2 = \frac{1}{\tilde{n}}\left(\sum_{t=a_i+1}^{a_i+\tilde{n}} Z_t\right)^2,
\]

where $\mathcal{A}$ is used as a placeholder to refer to any of the three types of region such that $\mathcal{A}_i  = \{a_i+1,\ldots, a_i + \tilde{n}\}$. Thus, the negative of the expression of interest is distributed according to $\chi^2_1$. Therefore, for sufficiently large $n$, the probability that this quantity is greater than $4\log \log n$ tends to $0$.

So we need focus only on the case where $k > 0$. Label, without loss of generality, the estimated changes which lie in the region $\mathcal{A}_i$ as $\hat{\tau}_1, \ldots, \hat{\tau}_k$, and let:

\[
\mbox{Diff} = \mbox{RSS}(y_{\mathcal{A}_i}; \hat{\tau}_{1:\hat{m}}) - \sum_{t=a_i+1}^{a_i + \tilde{n}} Z_t^2.
\]

Then:

\[
\mbox{Diff} = \frac{1}{\hat{\tau}_1 - a_i}\left(\sum_{t = a_i + 1}^{\hat{\tau}_1} Z_t\right)^2 + \ldots + \frac{1}{a_i+\tilde{n} - \hat{\tau}_k}\left(\sum_{t = \hat{\tau}_k + 1}^{a_i + \tilde{n}} Z_t\right)^2
\]

We demonstrate that this difference is less than $2k(1 + \epsilon)\log \tilde{n}$, for any $\epsilon > 0$. Note that, collectively, the positive terms in the expression follow a $\chi^2_{k+1}$ distribution. By \cite{LaurentMassart00}, for any quantity $U$ which follows a chi-squared distribution with $D$ degrees of freedom, then for any $x > 0$:

\begin{equation}\label{chi}
\mathbb{P}\left(U - D \geq 2 \sqrt{Dx} + 2 x\right) \leq \exp(-x).
\end{equation}

Letting $D = k + 1$ and $x = \frac{d \log \tilde{n} - \sqrt{\left(2d \log \tilde{n} - \left(k + 1\right)\right) \left(k + 1\right)}}{2}$, for some $d > 0$ such that $\tilde{n} \geq e^\frac{k+1}{2d}$. In practice $d>k$ (see below) so almost all positive integer values of $\tilde{n}$ will be sufficient. With this choice of $x$, the LHS of~(\ref{chi}) corresponds to $\mathbb{P}(U>d\log \tilde{n})$, and for large enough $\tilde{n}$~(\ref{chi}) becomes:

\begin{equation}\label{delta}
\mathbb{P}(U \geq d \log n) \leq \tilde{n}^{-\frac{d}{2} + \delta}, \text{ for any } \delta > 0
\end{equation}

There are then $\binom{\tilde{n}}{k}$ possible segmentations of these (incorrectly) fitted changes in this region. Given that $\binom{\tilde{n}}{k} < \frac{\tilde{n}^k}{k!}$ then by employing a Bonferroni correction, for the best segmentation involving $k$ changes in the region:

\begin{align*}
\mathbb{P}(\mbox{Diff} \geq d \log \tilde{n}) &\leq \tilde{n}^{-\frac{d}{2} + \delta} \tilde{n}^k \\
&= \tilde{n}^{k + \delta - \frac{d}{2}} \rightarrow 0 \text{ for } d=2k(1 + \epsilon), \text{ if we set, for example, } \delta = \epsilon/2.
\end{align*}

(For $d = 2k(1+ \epsilon)$, if $\delta = \epsilon/2$ - as (\ref{delta}) permits any strictly positive value of $\delta$ - then $k + \delta - \frac{d}{2} = -\left(2k - 1\right) \epsilon/2 < 0$.)

Note that this establishes the appropriate bound only in the case where $k$ is fixed and positive. To obtain the uniform bound over all k, we must sum over all $k = 1, ..., \tilde{n}$.

So for a given $\tilde{n}$ and $\epsilon$:

\begin{align*}
\sum_{k=1}^{\tilde{n}} \mathbb{P}(\mbox{Diff} \geq 2k \left(1+ \epsilon\right) \log \tilde{n}) &\leq \sum_{k=1}^{\tilde{n}} \tilde{n}^{-\left(2k -1\right)\epsilon/2} \\
&=\frac{\tilde{n}^{-\epsilon/2}\left(1 - \tilde{n}^{-\tilde{n}\epsilon}\right)}{1 - \tilde{n}^{-\epsilon}} \rightarrow 0, \text{ } \forall \epsilon > 0.
\end{align*}

This establishes the required results for both regions of type $\mathcal{B}_i$ and $\mathcal{L}_i$ ($\mathcal{R}_i$) by substituting $\tilde{n} = \lambda n$, $\lambda \leq 1$ and $\tilde{n} = \left\lceil\left(\log n\right)^{1 + \alpha}\right\rceil$ (for $\alpha < 1$ to obtain the constant $4$ in the two initial statements) respectively.

Hence $\mathbb{P}(\hat{m} > m) \rightarrow 0$.

\textbf{Proof of (b)}: Now have that $\hat{m} < m$. For $n$ sufficiently large, it is guaranteed that there is at least one true change (which shall be labelled $\tau$) such that the closest estimated change is at least $\left\lceil \left(\log n\right)^{1 + \alpha} \right\rceil$ time points away. Thus, by the proof of Corollary~\ref{errorval}, given that a change has been missed by this error, adding in estimated changes to the model at the points $\tau - \left\lceil \left(\log n\right)^{1 + \alpha} \right\rceil$, $\tau$, $\tau + \left\lceil \left(\log n\right)^{1 + \alpha} \right\rceil$ gives that the reduction in the RSS is greater than the incurred penalty for adding 3 changes. Thus, the original segmentation was not optimal.

Hence $\mathbb{P}(\hat{m} < m) \rightarrow 0$.

Lastly, we need to establish that when $\hat{m} = m$, the event that each of the estimated changes is within $\left\lceil \left(\log n\right)^{1 + \alpha} \right\rceil$ of a true change tends to 1. Suppose we have a segmentation with $\hat{m} = m$ which contains a true change, $\tau_i$, with no estimated changes within $\left\lceil \left(\log n\right)^{1 + \alpha} \right\rceil$. Then by comparing this segmentation to an equivalent segmentation which also fits estimated changes at $\tau_i - \left\lceil \left(\log n\right)^{1 + \alpha} \right\rceil, \tau_i, \tau_i + \left\lceil \left(\log n\right)^{1 + \alpha} \right\rceil$, we again obtain a saving of greater than the cost of adding 3 changes by~\cite{Yao88} and Corollary~\ref{errorval}. \QEDB

Note that this result extends naturally to a multivariate analogue:

\begin{lemma}\label{multiconsistency}
Take a procedure which exactly minimises the squared error loss for the multivariate problem:

\begin{equation}\label{multichangeinmean} 
\textbf{Y}_i = \boldsymbol{\epsilon}_i + \boldsymbol{\mu}_k,  \text{ for $\tau_{k-1} + 1 \leq i \leq \tau_k$,  and $k \in \{1, ..., m + 1\}$},
\end{equation}

where $\textbf{Y}_i = \left(Y_i^{\left(1\right)}, ..., Y_i^{\left(d\right)}\right)^T$, $\forall i \in \{1, ..., n\}$; $\boldsymbol{\mu}_k \neq \boldsymbol{\mu}_{k+1}$, $\forall k \in \{1,...,m\}$; $\boldsymbol{\epsilon}_i \sim^{i.i.d.} N_d\left(\textbf{0},\sigma^2 I\right)$, some $d$. In addition, take the penalty for fitting a change to be $\left(d + 1 \right)\left(1+\epsilon\right) \log n$, for any $\epsilon > 0$. Then defining $\mathcal{E}_n^{\alpha}$ as for Lemma~\ref{peltconsistency} for any $\alpha > 0$ again gives that $\mathbb{P}(\mathcal{E}_n^{\alpha}) \rightarrow 1$ as $n \rightarrow \infty$.
\end{lemma}

\textbf{Proof of Lemma~\ref{multiconsistency}}: We define the natural extension of the residual sum of squares in the multivariate case as:

\begin{align*}
\mbox{RSS}(\textbf{y}_{1:n}; \hat{\tau}_{1:{\hat{m}}}) &= \sum_{i=1}^{\hat{\tau}_1} \left(\textbf{y}_i - \hat{\boldsymbol{\mu}}_1\right)^{T}\left(\textbf{y}_i - \hat{\boldsymbol{\mu}}_1\right) + ... + \sum_{i = \hat{\tau}_{\hat{m}} + 1}^n \left(\textbf{y}_i - \hat{\boldsymbol{\mu}}_{\hat{m}+1}\right)^{T}\left(\textbf{y}_i - \hat{\boldsymbol{\mu}}_{\hat{m}+1}\right) \\
&=\sum_{j=1}^{\hat{m}+1}\sum_{i = \hat{\tau}_{j-1} + 1}^{\hat{\tau}_j} \sum_{k=1}^d \left(y_{i,k} - \hat{\mu}_{j,k}\right)^2, \text{ with } \hat{\mu}_{j,k} = \frac{1}{\hat{\tau}_j - \hat{\tau}_{j-1}} \sum_{i=\hat{\tau}_{j-1} + 1}^{\hat{\tau}_j} y_{i,k} = \bar{y}_{j,k}.
\end{align*}

Using this, we proceed along the same trajectory as for the previous proof. Suppose that $\hat{m}$ changes are detected by the procedure. Then we first show that:

\textbf{(a)}: $\mathbb{P}(\hat{m} > m) \rightarrow 0$,

\textbf{(b)}: $\mathbb{P}(\hat{m} < m) \rightarrow 0$.

\textbf{Proof of (a)}: Again let $c_n^{\left(\alpha\right)} = \left\lceil\left(\log n\right)^{1 + \alpha}\right\rceil$. Note first that an equivalent result to Corollary~\ref{errorval} holds in the multivariate case as the residual sum of squares between the points $\tau_i - c_n^{\left(\alpha\right)}$ and $\tau_i + c_n^{\left(\alpha\right)}$ (where $\tau_i$ is some true change missed by the procedure as before) satisfies:

\begin{align*}
\frac{\mbox{RSS}\left(\textbf{y}_{\tau_i - c_n^{\left(\alpha\right)} + 1:\tau_i + c_n^{\left(\alpha\right)}}; \emptyset\right) - \sum\limits_{k=1}^d \sum\limits_{j = \tau_i - c_n^{\left(\alpha\right)} + 1}^{\tau_i + c_n^{\left(\alpha\right)}} Z_{j,k}^2}{c_n^{\left(\alpha\right)}} &= \sum_{k=1}^d \frac{\left(\mu_k^{\left(i\right)} - \mu_k^{\left(i+1\right)}\right)^2}{2} - \frac{1}{2c_n^{\left(\alpha\right)}} \sum_{k=1}^d \left(\sum_j Z_{j,k}\right)^2 \\
&\hspace{7pt}+ \frac{D_k}{c_n^{\left(\alpha\right)}} \\
&\rightarrow \sum_{k=1}^d \frac{\left(\mu_k^{\left(i\right)}-\mu_k^{\left(i+1\right)}\right)^2}{2} \text{ as } n \rightarrow \infty,
\end{align*}

where $D_k$ is normally distributed with a variance equivalent to the deterministic term scaled by $4 \sigma^2$.

Hence, as per the previous proof, we can compare the residual sum of squares of the fit of a set of estimated changes with $\hat{m} > m$ across (equivalent) regions $\mathcal{B}_i, \mathcal{L}_i, \mathcal{R}_i$ to the null fit. Across a region bounded by the points $\left(a, b\right)$ containing estimated changes $\hat{\tau}_1, ..., \hat{\tau}_{p}$, the relevant difference term is:

\begin{align*}
\mbox{Diff} &= \sum\limits_{k=1}^d \left[\frac{1}{\hat{\tau}_1 - a}\left(\sum\limits_{j = a + 1}^{\hat{\tau}_1} Z_{j,k} \right)^2 + ... + \frac{1}{b - \hat{\tau}_p} \left(\sum\limits_{j = \hat{\tau}_p + 1}^b Z_{j,k} \right)^2\right],
\end{align*}

giving that $\mbox{Diff} \sim \chi^2_{d\left(p+1\right)}$. A similar argument to before then gives that:

\begin{align*}
\mathbb{P}(\mbox{Diff} \geq p\left(d + 1\right)\left(1+\epsilon\right) \log n) \rightarrow 0,
\end{align*}

and in particular:

\begin{align*}
\sum_{p=1}^n\mathbb{P}(\mbox{Diff} \geq p \left(d + 1\right)\left(1 + \epsilon\right) \log  n) \rightarrow 0.
\end{align*}

Hence $\mathbb{P}(\hat{m} > m) \rightarrow 0$.

\textbf{Proof of (b):} This follows immediately from considering the multivariate equivalent to Corollary \ref{errorval} shown above, inferring the presence of a missed change, $\tau_i$, and fitting three estimated changes at $\tau_i - \left\lceil\left(\log n\right)^{1 + \alpha}\right\rceil, \tau_i, \tau_i +\left\lceil \left(\log n\right)^{1+ \alpha} \right\rceil$. This segmentation will produce a lower residual sum of squares than the original with probability approaching 1.

Hence $\mathbb{P}(\hat{m} < m) \rightarrow 0$.

All that remains is to show that this correct number of changes falls within $\left\lceil\left(\log n\right)^{1 + \alpha} \right\rceil$. However, this again follows the same line of reason as for the univariate case by the result established above. \QEDB

\bibliographystyle{apa}

\end{document}